\documentclass[aps,preprint,prd]{revtex4.1}
\usepackage{graphicx}
\usepackage[bf,SL,BF]{subfigure}
\usepackage{psfrag}
\usepackage{color}
\usepackage{amssymb}
\usepackage{amsmath}
\usepackage{epstopdf}
\usepackage{hyperref}

\newcommand{\be}{\begin{eqnarray}}
\newcommand{\ee}{\end{eqnarray}}

\newcommand{\bfb}{{\bf b}_{\perp}}

\newcommand{\bfp}{{\bf p}_{\perp}}

\newcommand{\Dp}{{\bf \Delta}_{\perp}}

\begin{document}

\title{Sivers and Boer-Mulders GTMDs in Light-front Holographic Quark-diquark Model}
\author{Dipankar Chakrabarti$ ^1$, Narinder Kumar$^{1,2}$, Tanmay Maji$^{3,4}$, Asmita Mukherjee$ ^4$ }
\affiliation{ $ ^1$Department of Physics, 
Indian Institute of Technology Kanpur,
Kanpur 208016, India\\ 
$ ^2$Department of Physics, Doaba College Jalandhar, Jalandhar-144004, India\\
$^3$Key Laboratory of Nuclear Physics and Ion-beam Application (MOE) and\\ Institute of Modern Physics, Fudan University, Shanghai 200433, China\\  
$ ^4$Department of Physics, 
Indian Institute of Technology Bombay, Powai,
Mumbai 400076, India
}
\date{\today}
\begin{abstract}
We calculate the time reversal odd (T-odd) generalized transverse momentum dependent parton distributions (GTMDs) $F^o_{1,2}$ and $H^o_{1,1}$ in a light-front quark-diquark model based on ADS/QCD. In the limit of zero momentum transfer, these reduce to the Sivers and Boer-Mulders TMDs respectively.  We have incorporated both scalar and axial vector diquarks in the model and obtained an overlap representation of the GTMDs using the light-front wave functions (LFWFs). Contribution from the final state interaction is incorporated at the level of one gluon exchange as a phase factor in the LFWF. We show that the final state interaction term can be factored out in this model and this part is the same for both GTMDs. We also present the corresponding Wigner distributions.   
\end{abstract}
\pacs{14.20.Dh, 12.39.-x,12.38.Aw, 12.90.+b}
\maketitle
\section{Introduction\label{intro}}
In recent years, the generalized transverse momentum dependent parton
distributions (GTMDs) and Wigner distributions have gained quite a lot of
attention in the field of hadron physics. Off-forward parton correlators were 
introduced in \cite{Hagler:2003jw} in the context of the orbital angular momentum of the quarks. 
Generalized parton correlation functions (GPCFs) were introduced in \cite{Meissner:2009ww}
which are the fully unintegrated off-diagonal quark-quark
correlators. The GPCFs depend on the four momentum of the quark and on the
momentum transfer to the hadron. GTMDs are obtained from the GPCFs by
integrating over the light cone component of the quark momentum. These are
the so-called ``mother distributions''.  The generalized parton distributions
(GPDs) and transverse momentum dependent parton distributions (TMDs) are
obtained by taking certain limits of the GTMDs \cite{Lorce:2013pza}. A Fourier transform of the
GTMDs with respect to the momentum transfer $ \Delta_\perp$ gives the Wigner
distributions \cite{Wigner:1932eb}, which are quantum mechanical analogs of the classical phase
space distribution  of quarks and gluons in the nucleon. Wigner
distributions had been used in different branches of physics for example
in quantum information, optics, image processing, quark-gluon plasma,
nonlinear dynamics etc. The Wigner distributions for quarks and gluons
were  first introduced in \cite{Ji:2003ak, Belitsky:2003nz}  as a six dimensional object (three position
and three momentum coordinates) in the rest frame of the nucleon. The five
dimensional Wigner distributions in the infinite momentum frame or
equivalently in the light cone framework was introduced in \cite{Lorce:2011kd}. A positive definite Husimi distribution was introduced in \cite{Hagiwara:2014iya,Hatta:2015ggc} to study nucleon tomography. GTMDs and
Wigner distributions are also shown to be related to the elusive quark and
gluon orbital angular momentum and spin-orbit correlations \cite{Lorce:2011ni, Hatta:2011ku, Lorce:2011ni}.  Although
experimental determination of Wigner distribution is challenging, there are
already several theoretical studies and model calculations, for
example, in constituent quark model and chiral quark soliton model 
\cite{Lorce:2011kd},
spectator model \cite{Liu:2015eqa}, color-glass condenstate \cite{Hagiwara:2016kam}, and 
also in a holographic model  \cite{Chakrabarti:2017teq}.  Wigner distributions for both quarks and gluons in a perturbative dressed quark model were calculated in \cite{Mukherjee:2014nya, Mukherjee:2015aja, More:2017zqq, More:2017zqp}.
 In \cite{zhou:2016rnt} 
Wigner distribution for a nucleus was presented.  The multipole decomposition of the nucleon in transverse space was discussed in \cite{Lorce:2015sqe}. Evolution of the GTMDs was studied in \cite{Echevarria:2016mrc}.  The recent paper
\cite{Ma:2018ysi} presents a calculation of the Wigner distribution for pions using light-cone wave
functions. The final state interaction is important in particular for the
time-reversal odd (T-odd) GTMDs  and Wigner functions. These interactions are
essential for generating the single spin asymmetries \cite{Brodsky:2002cx}  for example the Sivers
asymmetry. These are in fact contribution of the gauge link or Wilson line
present in the correlator for color gauge invariance.   In \cite{Brodsky:2002cx} it was first shown in a model
calculation with one gluon exchange to generate the necessary phase factor for Sivers asymmetry. Later such approach has been used in similar models to introduce a LFWF with the phase included to calculate single spin asymmetries  \cite{Brodsky:2010vs}.  In \cite{Kumar:2015coa}, Sivers  asymmetries 
were evaluated in a spin-1 diquark model and in  \cite{Maji:2017wwd}, we have calculated Sivers and $cos ~2 \phi$ asymmetries in a diquark model that includes both scalar and axial vector diquarks.  For the GTMDs, most model
calculations so far do not include the final and initial state interactions. In this work, for the first time, we present a calculation of the T-odd Sivers and Boer-Mulders GTMDs, including the final state
interaction. We use a light-front spectator diquark model, with scalar and
vector diquarks. The light-front wave functions (LFWFs) are predicted by the
soft-wall ADS/QCD. The model is extended for the T-odd GTMDs to include the
final state interaction upto one gluon exchange in the LFWF. It is known that in spectator type
models, it is possible to express the T-odd TMDs in terms of a final state interaction function
multiplied with a term that is independent of this interaction \cite{Bacchetta:2008af}. 
We observe that also in our model, a final state interaction function can be factored out from 
the T-odd GTMDs. We also calculate the Wigner distributions corresponding to 
these GTMDs by taking a Fourier transform  with respect to the transverse momentum transfer. 

The plan of the paper is as follows. In section \ref{model} we present the light-front diquark model. Final state interactions 
and T-odd GTMDs are discussed in sections \ref{Sec_FSI} and \ref{Todd} respectively. Wigner functions are presented in section \ref{wigner}. Numerical results are  given in    section \ref{results}.  Discussions about the final state interaction function in this model are  given in section \ref{fsi}.  Section \ref{summary} gives the summary and conclusion.
\section{light-front quark-diquark model for the nucleon \label{model}}
To investigate the T-odd GTMDs, we consider a light-front quark-diquark model 
for the proton. In this model, the proton state is  considered to be a linear 
combination of quark-diquark state including both the scalar and axial-vector 
diquarks with $SU(4)$  spin-flavor structure and can be written 
as \cite{Jakob:1997wg, Bacchetta:2008af, Maji:2016yqo}
\be 
|P; \pm\rangle = C_S|u~ S^0\rangle^\pm + C_V|u~ A^0\rangle^\pm + C_{VV}|d~ A^1\rangle^\pm. \label{PS_state}
\ee
Where, $C_S, C_V$ and $C_{VV}$ are the coefficient of the isoscalar-scalar 
diquark singlet state $|u~ S^0\rangle$, isoscalar-axial vector diquark state 
$|u~ A^0\rangle$ and isovector-axial vector diquark state $|d~ A^1\rangle$ 
respectively. $S$ and $A$ represent the scalar and axial-vector diquark with 
isospin at their superscript. Under the isospin symmetry, the neutron  state is 
defined by the above formula with $u\leftrightarrow d$.

The two particle Fock-state expansion for $J^z =\pm1/2$ for spin-0 diquark state 
is given by
\be
|u~ S\rangle^\pm & =& \int \frac{dx~ d^2\bfp}{2(2\pi)^3\sqrt{x(1-x)}} \bigg[ \psi^{\pm(u)}_{+}(x,\bfp)|+\frac{1}{2}~s; xP^+,\bfp\rangle \nonumber \\
 &+& \psi^{\pm(u)}_{-}(x,\bfp)|-\frac{1}{2}~s; xP^+,\bfp\rangle\bigg],\label{fock_PS}
\ee
where $|\lambda_q~\lambda_S; xP^+,\bfp\rangle$ is the two particle state having 
struck quark of helicity $\lambda_q$ and a scalar diquark having helicity 
$\lambda_S=s$. 
The state with spin-1 diquark is given as \cite{Ellis:2008in}
\be
|\nu~ A \rangle^\pm & =& \int \frac{dx~ d^2\bfp}{2(2\pi)^3\sqrt{x(1-x)}} \bigg[ \psi^{\pm(\nu)}_{++}(x,\bfp)|+\frac{1}{2}~+1; xP^+,\bfp\rangle \nonumber\\
 &+& \psi^{\pm(\nu)}_{-+}(x,\bfp)|-\frac{1}{2}~+1; xP^+,\bfp\rangle +\psi^{\pm(\nu)}_{+0}(x,\bfp)|+\frac{1}{2}~0; xP^+,\bfp\rangle \nonumber \\
 &+& \psi^{\pm(\nu)}_{-0}(x,\bfp)|-\frac{1}{2}~0; xP^+,\bfp\rangle + \psi^{\pm(\nu)}_{+-}(x,\bfp)|+\frac{1}{2}~-1; xP^+,\bfp\rangle \nonumber\\
 &+& \psi^{\pm(\nu)}_{--}(x,\bfp)|-\frac{1}{2}~-1; xP^+,\bfp\rangle  \bigg].\label{fock_PS}
\ee

\begin{figure}
 \includegraphics[width=10cm,clip]{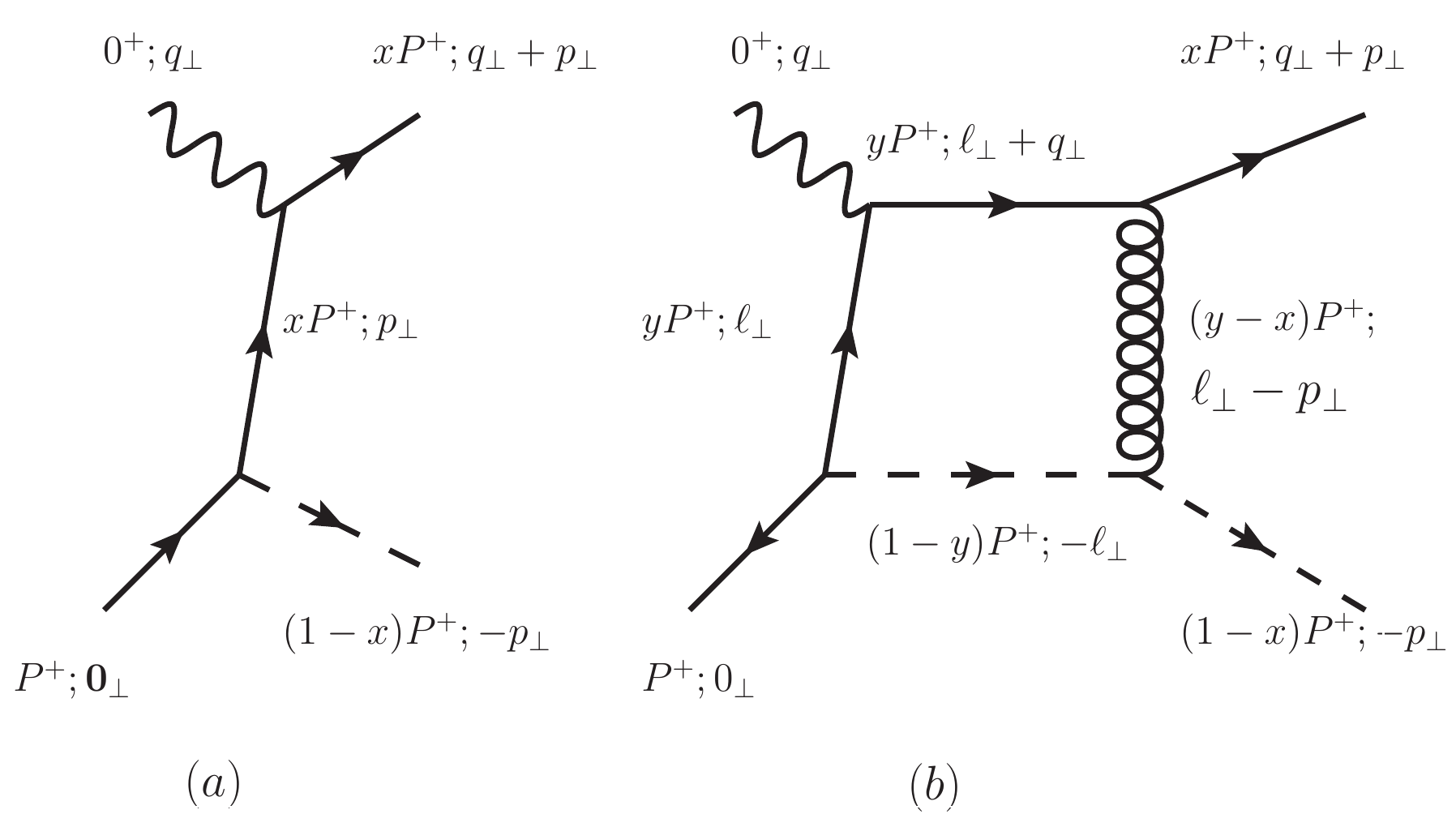}
 \caption{\label{fig_FSI} Left: tree level diagram. Right: FSI diagram for 
$\gamma^* P \to q(qq)$ }
 \end{figure}

\section{Final State interaction and T-odd GTMDs \label{Sec_FSI}}
The light cone wave function representations of the T-odd TMDs and GTMDs 
requires that the wave functions must have complex phases. 
Brodsky, Hwang and Schmidt \cite{Brodsky:2002cx} showed that
the final state interaction produces a non-trivial phase
in the amplitude which generates the Sivers asymmetry. If we assume that the 
QCD factorization theorem holds for SIDIS and DY processes, Sivers and Boer 
Mulders asymmetries can be written as convolutions of T-odd TMDs (Sivers and 
Boer-Mulders functions), a hard part and the fragmentation function. 
To have a wave function representation of the T-odd TMDs and GTMDs, we 
incorporate the final state interaction into the light-cone wave functions  with 
complex phases. The modified wave functions give 
 non-vanishing T-odd GTMDs along with the T-even GTMDs. 
The modified wave functions \cite{Hwang:2010dd} are written in the following forms:\\
(i) for scalar diquark
\be 
\psi^{+(u)}_+(x,\bfp)&=& N_S~\bigg[1+i \frac{e_1 e_2}{8 \pi}(\bfp^2 + B)g_1 \bigg] \varphi^{(u)}_{1}(x,\bfp),\nonumber \\
\psi^{+(u)}_-(x,\bfp)&=& N_S\bigg(- \frac{p^1+ip^2}{xM} \bigg) \bigg[1+i \frac{e_1 e_2}{8 \pi}(\bfp^2 + B)g_2\bigg]  \varphi^{(u)}_{2}(x,\bfp),\label{LFWF_S}\\
\psi^{-(u)}_+(x,\bfp)&=& N_S \bigg(\frac{p^1-ip^2}{xM}\bigg) \bigg[1+i \frac{e_1 e_2}{8 \pi}(\bfp^2 + B)g_2\bigg] \varphi^{(u)}_{2}(x,\bfp),\nonumber \\
\psi^{-(u)}_-(x,\bfp)&=&  N_S~ \bigg[1+i \frac{e_1 e_2}{8 \pi}(\bfp^2 + B)g_1\bigg]\varphi^{(u)}_{1}(x,\bfp),\nonumber
\ee
 (ii) for axial-vector diquark (for $J= +1/2$ ) 
\be 
\psi^{+(\nu)}_{+~+}(x,\bfp)&=& N^{(\nu)}_1 \sqrt{\frac{2}{3}} \bigg(\frac{p^1-ip^2}{xM}\bigg)\bigg[1+i \frac{e_1 e_2}{8 \pi}(\bfp^2 + B)g_2 \bigg] \varphi^{(\nu)}_{2}(x,\bfp),\nonumber \\
\psi^{+(\nu)}_{-~+}(x,\bfp)&=& N^{(\nu)}_1 \sqrt{\frac{2}{3}} \bigg[1+i \frac{e_1 e_2}{8 \pi}(\bfp^2 + B)g_1 \bigg] \varphi^{(\nu)}_{1}(x,\bfp),\nonumber \\
\psi^{+(\nu)}_{+~0}(x,\bfp)&=& - N^{(\nu)}_0 \sqrt{\frac{1}{3}} \bigg[1+i \frac{e_1 e_2}{8 \pi}(\bfp^2 + B)g_1 \bigg] \varphi^{(\nu)}_{1}(x,\bfp),\label{LFWF_Vp}\\
\psi^{+(\nu)}_{-~0}(x,\bfp)&=& N^{(\nu)}_0 \sqrt{\frac{1}{3}} \bigg(\frac{p^1+ip^2}{xM} \bigg) \bigg[1+i \frac{e_1 e_2}{8 \pi}(\bfp^2 + B)g_2 \bigg] \varphi^{(\nu)}_{2}(x,\bfp),\nonumber \\
\psi^{+(\nu)}_{+~-}(x,\bfp)&=& 0,\nonumber \\
\psi^{+(\nu)}_{-~-}(x,\bfp)&=&  0, \nonumber 
\ee
and for $J=-1/2$
\be 
\psi^{-(\nu)}_{+~+}(x,\bfp)&=& 0,\nonumber \\
\psi^{-(\nu)}_{-~+}(x,\bfp)&=& 0,\nonumber \\
\psi^{-(\nu)}_{+~0}(x,\bfp)&=& N^{(\nu)}_0 \sqrt{\frac{1}{3}} \bigg( \frac{p^1-ip^2}{xM} \bigg) \bigg[1+i \frac{e_1 e_2}{8 \pi}(\bfp^2 + B)g_2 \bigg] \varphi^{(\nu)}_{2}(x,\bfp),\label{LFWF_Vm}\\
\psi^{-(\nu)}_{-~0}(x,\bfp)&=& N^{(\nu)}_0\sqrt{\frac{1}{3}} \bigg[1+i \frac{e_1 e_2}{8 \pi}(\bfp^2 + B)g_1 \bigg] \varphi^{(\nu)}_{1}(x,\bfp), \nonumber \\
\psi^{-(\nu)}_{+~-}(x,\bfp)&=& - N^{(\nu)}_1 \sqrt{\frac{2}{3}} \bigg[1+i \frac{e_1 e_2}{8 \pi}(\bfp^2 + B)g_1 \bigg] \varphi^{(\nu)}_{1}(x,\bfp), \nonumber \\
\psi^{-(\nu)}_{-~-}(x,\bfp)&=& N^{(\nu)}_1 \sqrt{\frac{2}{3}} \bigg(\frac{p^1+ip^2}{xM}\bigg) \bigg[1+i \frac{e_1 e_2}{8 \pi}(\bfp^2 + B)g_2 \bigg] \varphi^{(\nu)}_{2}(x,\bfp),\nonumber
\ee
where,
\be 
g_1 &=& \int^1_0 d\alpha \frac{-1}{\alpha(1-\alpha)\bfp^2 + \alpha m_g^2 + 
(1-\alpha)B} ,\label{g1}\\
g_2 &=& \int^1_0 d\alpha \frac{-\alpha}{\alpha(1-\alpha)\bfp^2 + \alpha m_g^2 + 
(1-\alpha)B},\label{g2}\\
{\rm and},\nonumber\\
B &=& x(1-x)(-M^2+\frac{m^2_q}{x}+\frac{m^2_D}{1-x})\label{B}.
\ee
%
\be
\varphi_i^{(\nu)}(x,\bfp)&=& \frac{4\pi}{\kappa}\sqrt{\frac{\log(1/x)}{1-x}}x^{a_i^\nu}(1-x)^{b_i^\nu}\exp\bigg[-\delta^\nu\frac{\bfp^2}{2\kappa^2}\frac{\log(1/x)}{(1-x)^2}\bigg],
\label{LFWF_phi}\nonumber\\
&=& A^\nu_i(x) \exp\bigg[-a(x)\bfp^2 \bigg]
\ee
with
\be 
 A^\nu_i(x) &=& \frac{4\pi}{\kappa}\sqrt{\frac{\log(1/x)}{1-x}}x^{a_i^\nu}(1-x)^{b_i^\nu},\\
\tilde{a}(x)&=&\delta^\nu \frac{\log(1/x)}{2\kappa^2 (1-x)^2};~~~{\rm and}~~ 
a(x)= 2 \tilde{a}(x).
\ee
 The wave functions of Eqs.(\ref{LFWF_S}-\ref{LFWF_Vm}) carry nucleon helicity at the superscript and quark helicity at the subscript. $\nu$ is flavor index for the flavors $u$ and $d$. In Eq.(\ref{B}), proton mass, quark mass and the diquark mass are denoted by $M, m_q$ and $m_D$ respectively. $e_1$ and $e_2$ are the color charge of the struck quark and diquark respectively and FSI gauge exchange strength is $\frac{e_1 e_2}{4\pi}$. The parameters $a^{\nu}_i, b^\mu_i$ and $\delta^\nu$ are the same as introduced in \cite{Maji:2016yqo,Maji:2017bcz} where, the parameters are determined from the fitting of the Dirac and Pauli form factors data. The normalization constants $N_i's~(i=S,0,1)$ are fixed by the quark counting rule for proton \cite{Maji:2016yqo,Maji:2017bcz}.Note that here the FSI contribution comes as a phase factor to the wave functions and that phase factor do not contribute to the T-even observables and the same parameters can be used.  
%

\section{T-odd GTMDs}\label{Todd}
The GTMD correlator is defined as
\be
W^{\nu [\Gamma]}_{\lambda'' \lambda'}(\Dp,\bfp,x)=\frac{1}{2}\int \frac{dz^-}{(2\pi)} \frac{d^2z_T}{(2\pi)^2} e^{ip.z} 
\langle P^{\prime\prime}; \lambda''|\bar{\psi}^\nu _i(-z/2)\Gamma \mathcal{W}_{[-z/2,z/2]} \psi^\nu _j(z/2) |P^\prime;\lambda'\rangle \bigg|_{z^+=0}
\label{GTMD_ll}
\ee
for the twist-two Dirac $\gamma$-matrix $\Gamma=\gamma^+$, $\gamma^+\gamma_5$ or $i\sigma^{j+}\gamma_5$ (with $j=1, 2$) corresponding to unpolarized, longitudinally polarized or $j$-direction transverse polarized quark respectively. $\mathcal{W}_{[-z/2,z/2]}$ represents the gauge link Wilson line. As discussed before, we have included the contribution from the Wilson line in the form of final state interaction in the LFWFs.  
$|P' \lambda'\rangle$ represents the initial state of proton with momentum $P'$ and helicity $\lambda'$ and $|P'' \lambda''\rangle$ represents the final state of proton with momentum $P''$ and helicity $\lambda''$. In the symmetric frame, the kinematical variables are 
\be
P^\mu=\frac{(P'+P'')^\mu}{2}, \quad\quad \Delta^\mu=(P''-P')^\mu,
\ee
where $\Delta_\perp$ is the momentum transfer in this process.  We use the light-front coordinates $v^\mu=[v^+,v^-,\vec v_\perp]$, where $v^\pm=(v^0\pm v^3)$ and $\vec v_\perp=(v^1,v^2)$. 

In the bilinear decomposition, the correlator can be expressed in terms of different GTMDs for different polarization combination of proton and quarks \cite{Meissner:2009ww}. There are altogether 16 GTMDs at the leading twist and each GTMDs has an even and an odd part under the time reversal invariance. In this work, we will concentrate on two leading twist T-odd GTMDs which reduce to the Sivers TMD and Boer-Mulders TMD in the limit $\Delta_\perp=0$. At the TMD limit ($\Delta_\perp = 0$), the GTMDs $F^o_{1,2}$ and $H^o_{1,1}$ reduce to the Sivers function and Boer-Mulders function respectively and these GTMDs are defined as
\be 
W^{\nu [\gamma^+]}_{\lambda'' \lambda'}(\Dp,\bfp,x)&=& ... + \frac{1}{2M}\bar{u}(P'',\lambda'')\bigg(\frac{i \sigma^{i+}p^i_\perp}{P^+}\bigg)\bigg[F^{e\nu}_{1,2}+i F^{o \nu}_{1,2} \bigg]u(P',\lambda') + ...,\label{W_F12} \\
W^{\nu [i\sigma^{j+}\gamma^5]}_{\lambda'' \lambda'}(\Dp,\bfp,x)&=& \frac{1}{2M}\bar{u}(P'',\lambda'')(-\frac{i\epsilon^{ij}_T p^i_\perp}{M}) \bigg[H^{e\nu}_{1,1}+iH^{o\nu}_{1,1} \bigg]u(P',\lambda') + ..., \label{W_H11}
\ee
where the superscripts $e,o$ stand for T-even and T-odd part respectively.

\section{Wigner distributions}\label{wigner}
 In light-front framework, the 
5-dimensional quark Wigner distribution is defined as \cite{Lorce:2011kd,Lorce:2011ni}
\be
\rho^{\nu [\Gamma]}(\bfb,\bfp,x;S)=\int \frac{d^2\Dp}{(2\pi)^2} e^{-i\Dp.b_\perp} W^{\nu [\Gamma]}(\Dp,\bfp,x;S).
\label{wig_rho}
\ee
The correlator $W^{[\Gamma]}$ relates the GTMDs \cite{Meissner:2009ww} and in the Drell-Yan-West frame ($\Delta^+=0$) and fixed light-cone time $ z^+=0$ is given by
\be
W^{\nu [\Gamma]}(\Dp,\bfp,x;S)=\frac{1}{2}\int \frac{dz^-}{(2\pi)} \frac{d^2z_T}{(2\pi)^2} e^{ip.z} 
\langle P^{\prime\prime}; S|\bar{\psi}^\nu _i(z/2)\Gamma \mathcal{W}_{[-z/2,z/2]} \psi^\nu _j(z/2) |P^\prime;S\rangle \bigg|_{z^+=0}.
\label{wigner_W}
\ee
For a particular proton polarization $S$ the above correlator can be expressed as a linear combination of the helicity dependent GTMD correlator defined in Eq.(\ref{GTMD_ll}). Depending on the various polarization configurations of the proton ($X$) and the quark ($Y$), there are 16 independent twist-2 quark Wigner distributions ($\rho^\nu_{XY}$). In this section, we discuss $\rho^{i \nu }_{TU}$ and $\rho^{j \nu }_{UT}$ which are defined as
\be 
\rho^{i \nu }_{TU}(\bfb,\bfp,x)&=&\frac{1}{2}[\rho^{\nu [\gamma^+]}(\bfb,\bfp,x; +\hat{S}_i) - \rho^{\nu [\gamma^+]}(\bfb,\bfp,x; -\hat{S}_i)],\label{rho_TU_def}\\
\rho^{j \nu}_{UT}(\bfb,\bfp,x)&=&\frac{1}{2}[\rho^{\nu [i\sigma^{j+}\gamma^5]}(\bfb,\bfp,x; +\hat{S}_z) +\rho^{\nu [i\sigma^{j+}\gamma^5]}(\bfb,\bfp,x; -\hat{S}_z)]\label{rho_UT_def}.
\ee 
These two Wigner distributions can be parametrized in terms of the $F_{1,2}$ and $H_{1,1}$ GTMDs as
\be 
\rho^{i\nu}_{TU}(\bfb,\bfp,x)&=&\frac{1}{2M}\epsilon^{ij}_\perp \frac{\partial}{\partial b^j_\perp} \bigg[\mathcal{F}^\nu_{1,1}(x,0,\bfp^2,\bfp.\bfb,\bfb^2)-2\mathcal{F}^\nu_{1,3}(x,0,\bfp^2,\bfp.\bfb,\bfb^2)\bigg] \nonumber\\
&&+ i \frac{1}{M}\epsilon^{ij}_\perp p^j_\perp \mathcal{F}^\nu_{1,2}(x,0,\bfp^2,\bfp.\bfb,\bfb^2),\label{rhoTU_F}\\
\rho^{j \nu}_{UT}(\bfb,\bfp,x)&=& -i\frac{1}{M}\epsilon^{ij}_\perp p^i_\perp \mathcal{H}^\nu_{1,1}(x,0,\bfp^2,\bfp.\bfb,\bfb^2)\nonumber\\&&~~~~~~~~~~~+ \frac{1}{M}\epsilon^{ij}_\perp \frac{\partial}{\partial b^i_\perp} \mathcal{H}^\nu_{1,2}(x,0,\bfp^2,\bfp.\bfb,\bfb^2), \label{rhoUT_H}
\ee
Where the $\chi^\nu = \mathcal{F}^\nu_{1,1}, \mathcal{F}^\nu_{1,2},$ and $\mathcal{H}_{1,1},\mathcal{H}_{1,2}$ can be expressed as Fourier transform of GTMDs $X^\nu= F^\nu_{1,1}, F^\nu_{1,2}, F^\nu_{1,3}$ and $H_{1,1},H_{1,2}$ respectively.
\be 
\chi^\nu(x,0,\bfp^2,\bfp.\bfb,\bfb^2) = \int\frac{d^2\Dp}{(2\pi)^2} e^{-i\Dp.\bfb} X^\nu(x,0,\bfp^2,\bfp.\Dp,\Dp^2).\label{chi_GTMDs}
\ee
As we discussed before, each GTMD  can be written by separating the even (superscript $e$) and odd (superscript $o$) part under time-reversal as
\be
X^\nu(x,0,\bfp^2,\bfp.\Dp,\Dp^2) = X^{e\nu}(x,0,\bfp^2,\bfp.\Dp,\Dp^2)+i X^{o\nu}(x,0,\bfp^2,\bfp.\Dp,\Dp^2).
\ee
Here in this work, as we include explicate gluon in terms of final state interaction, GTMDs have non-vanishing odd part which contribute to the Wigner distributions \cite{Brodsky:2002cx,Lorce:2011ni}. The GTMDs are complex quantities because of the T-odd part coming from the FSI. However, the Hermiticity property of the GTMDs ensures that the Wigner distributions are real-valued functions. Hermiticity of the GTMDs is satisfied in our model and as a result, the phase space distributions are real.

Let us define the Fourier transform of these leading twist T-odd GTMDs as
\be 
\rho^\nu_{Siv}( \bfb, \bfp, x) &=& \int\frac{d^2\Dp}{(2\pi)^2} e^{-i\Dp.\bfb} F^{o\nu}_{1,2}(\Dp,\bfp,x),\label{RSiv}\\
\rho^\nu_{BM}( \bfb, \bfp, x) &=& \int\frac{d^2\Dp}{(2\pi)^2} e^{-i\Dp.\bfb}  H^{o\nu}_{1,1}(\Dp,\bfp,x), \label{RBM}
\ee
where, the superscript $``o"$ in $ F^{o\nu}_{1,2}$ and $H^{o\nu}_{1,1}$ represents the T-odd part of the respective GTMDs. Since, at the TMD limit $\Dp=0$, the associated GTMDs $ F^{o\nu}_{1,2}$ and $H^{o\nu}_{1,1}$ reduce to Sivers and Boer-Mulders TMDs, we refer these Wigner distributions by Sivers and Boer-Mulder Wigner distributions respectively and are labelled by subscripts $Siv$ and $BM$. 
 
The modified Wigner distribution  $\hat{\rho}^{i\nu}_{TU}( \bfb, \bfp, x)$ is given by
\be
\hat{\rho}^{i\nu}_{TU}(\bfb,\bfp,x)&=&\frac{1}{2M}\epsilon^{ij}_\perp \frac{\partial}{\partial b^j_\perp} \int\frac{d^2\Dp}{(2\pi)^2} e^{-i\Dp.\bfb} \bigg[ F^{e \nu}_{1,1}(\Dp, \bfp, x) - 2 F^{e \nu}_{1,3}(\Dp, \bfp, x) \bigg] \nonumber\\
&&- \frac{1}{M}\epsilon^{ij}_\perp p^j_\perp \int\frac{d^2\Dp}{(2\pi)^2} e^{-i\Dp.\bfb} F^{o\nu}_{1,2}(\Dp, \bfp, x),\\
&=& \rho^{i\nu}_{TU}(\bfb,\bfp,x)  - \frac{1}{M}\epsilon^{ij}_\perp p^j_\perp  \rho^\nu_{Siv}( \bfb, \bfp, x).\label{RUTSiv}
\ee
Similarly, $\hat{\rho}^{\nu}_{UT}( \bfb, \bfp, x)$ is given as
\be
\hat{\rho}^{j \nu}_{UT}(\bfb,\bfp,x)&=& \frac{1}{M}\epsilon^{ij}_\perp \frac{\partial}{\partial b^i_\perp} \int\frac{d^2\Dp}{(2\pi)^2} e^{-i\Dp.\bfb} H^{e\nu}_{1,2}(\Dp,\bfp,x)\nonumber \\
&& ~~~ +\frac{1}{M}\epsilon^{ij}_\perp p^i_\perp \int\frac{d^2\Dp}{(2\pi)^2} e^{-i\Dp.\bfb}  H^{o\nu}_{1,1}(\Dp,\bfp,x)\\
&=& \rho^{j \nu}_{UT}(\bfb,\bfp,x) +\frac{1}{M}\epsilon^{ij}_\perp p^i_\perp \rho^\nu_{BM}( \bfb, \bfp, x). \label{RUTBM}
\ee
In Eq.(\ref{RUTSiv}), superscripts $i$ represents the transverse polarization 
direction of proton and superscript $j$ in Eq.({\ref{RUTBM}) stands for the 
quark transverse polarization direction. The last terms in Eq. (\ref{RUTSiv}) and (\ref{RUTBM}) are the modifications of the Wigner distribution due to the FSI term. Analytical as well as numerical model results for $\hat{\rho}^{j \nu}_{TU}$ and $\hat{\rho}^{j \nu}_{UT}$ are discussed in the following section-\ref{results}.

\begin{figure}
\includegraphics[scale=0.38]{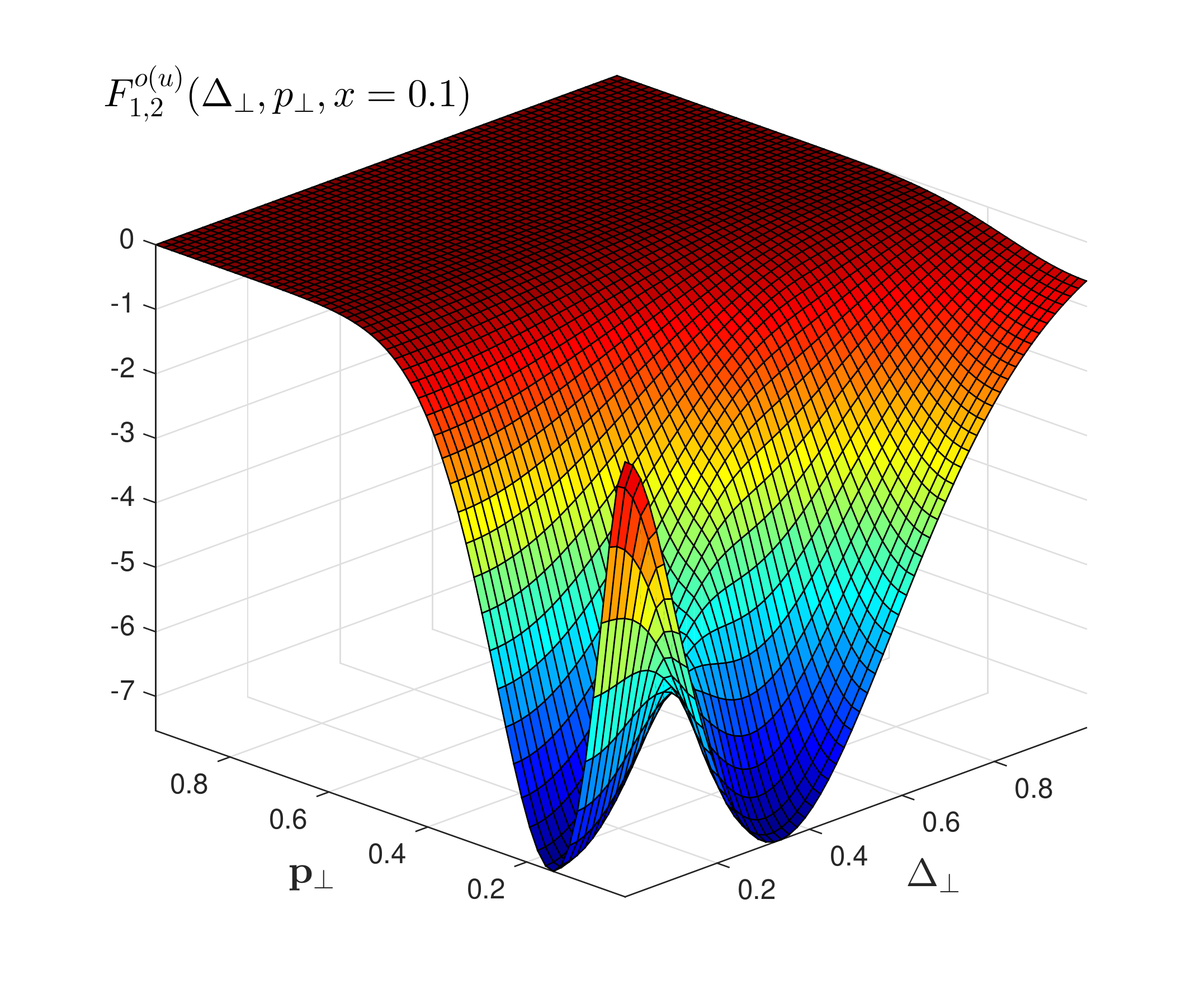} 
\includegraphics[scale=0.38]{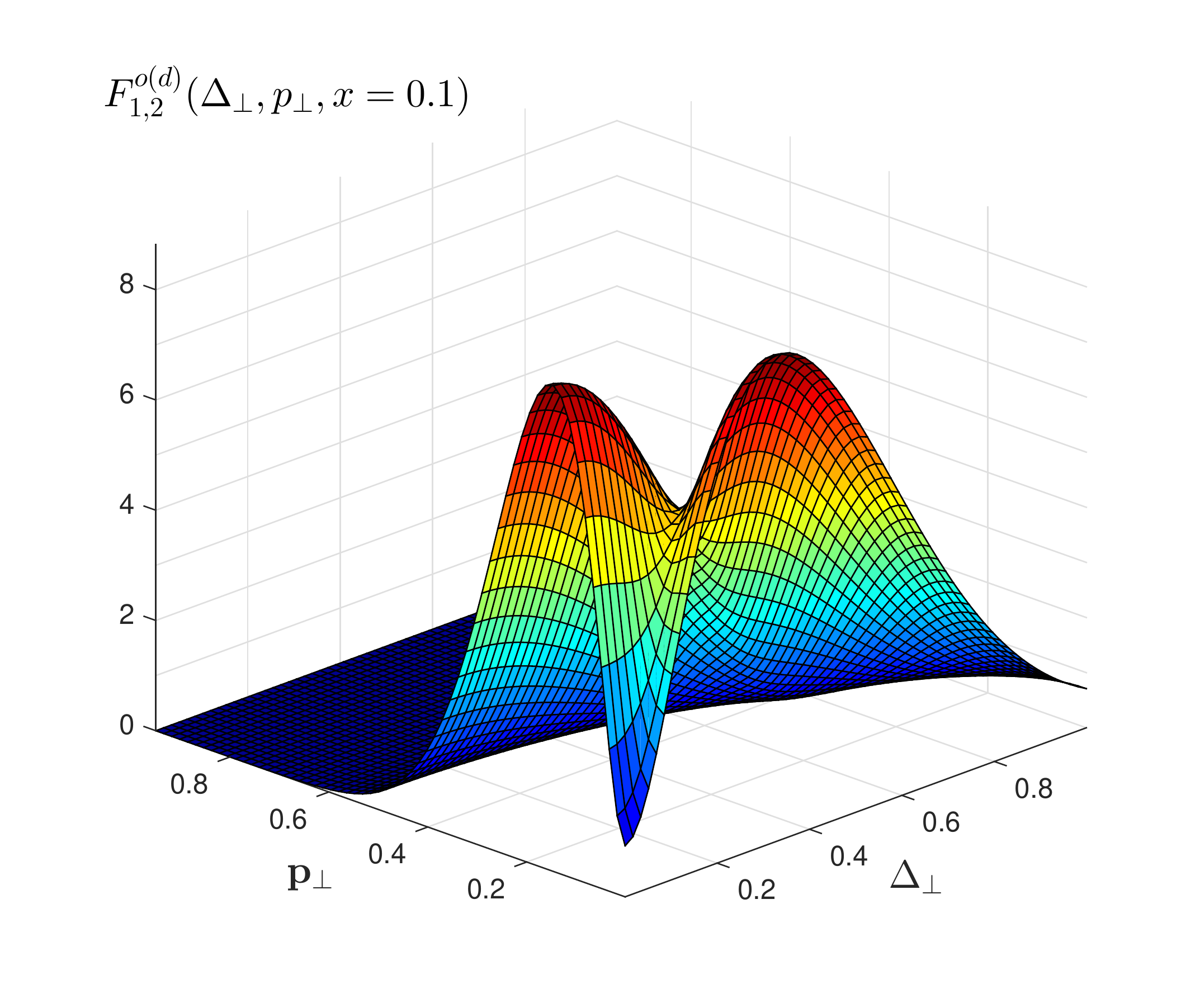} \\
\includegraphics[scale=0.38]{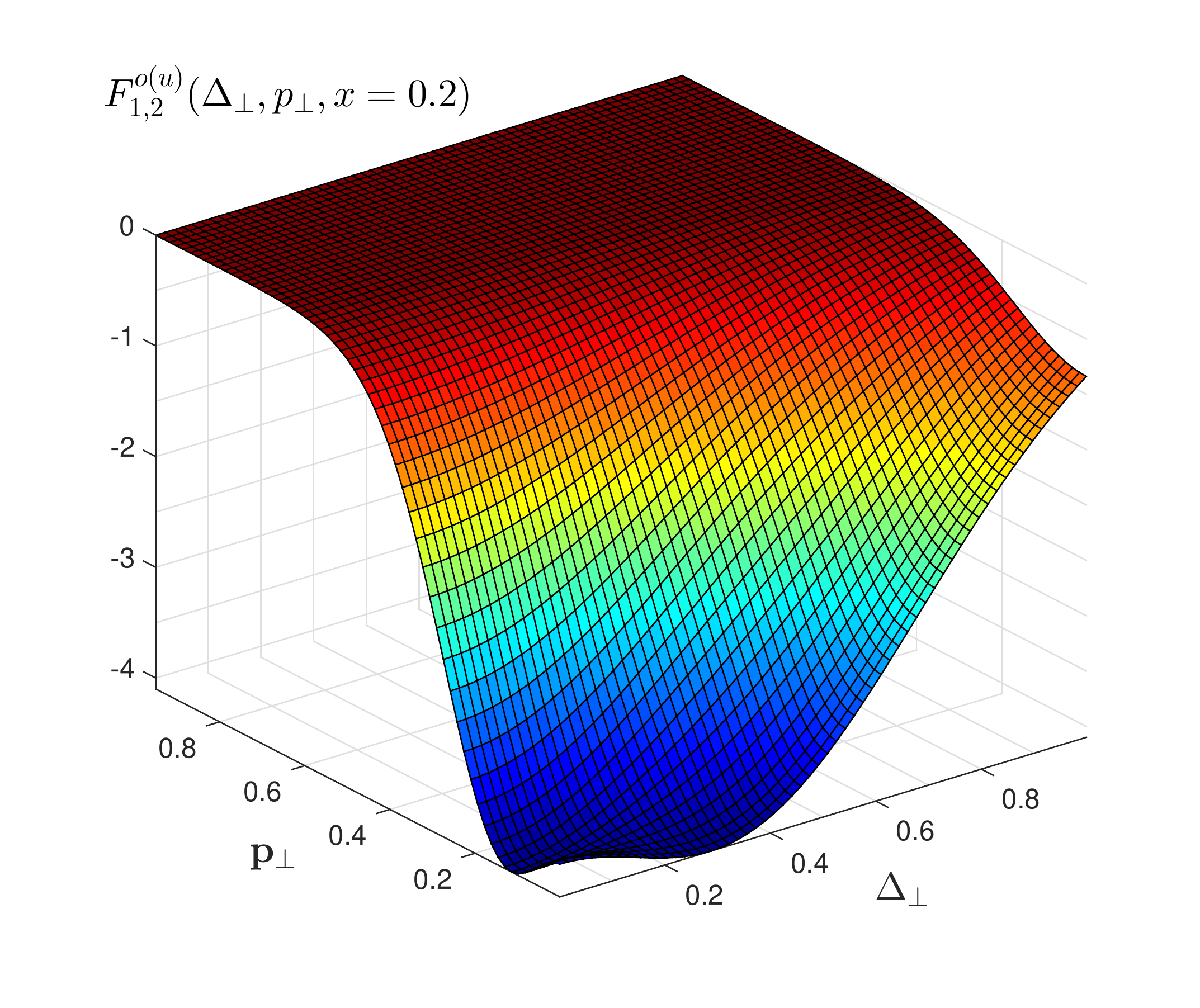}
\includegraphics[scale=0.38]{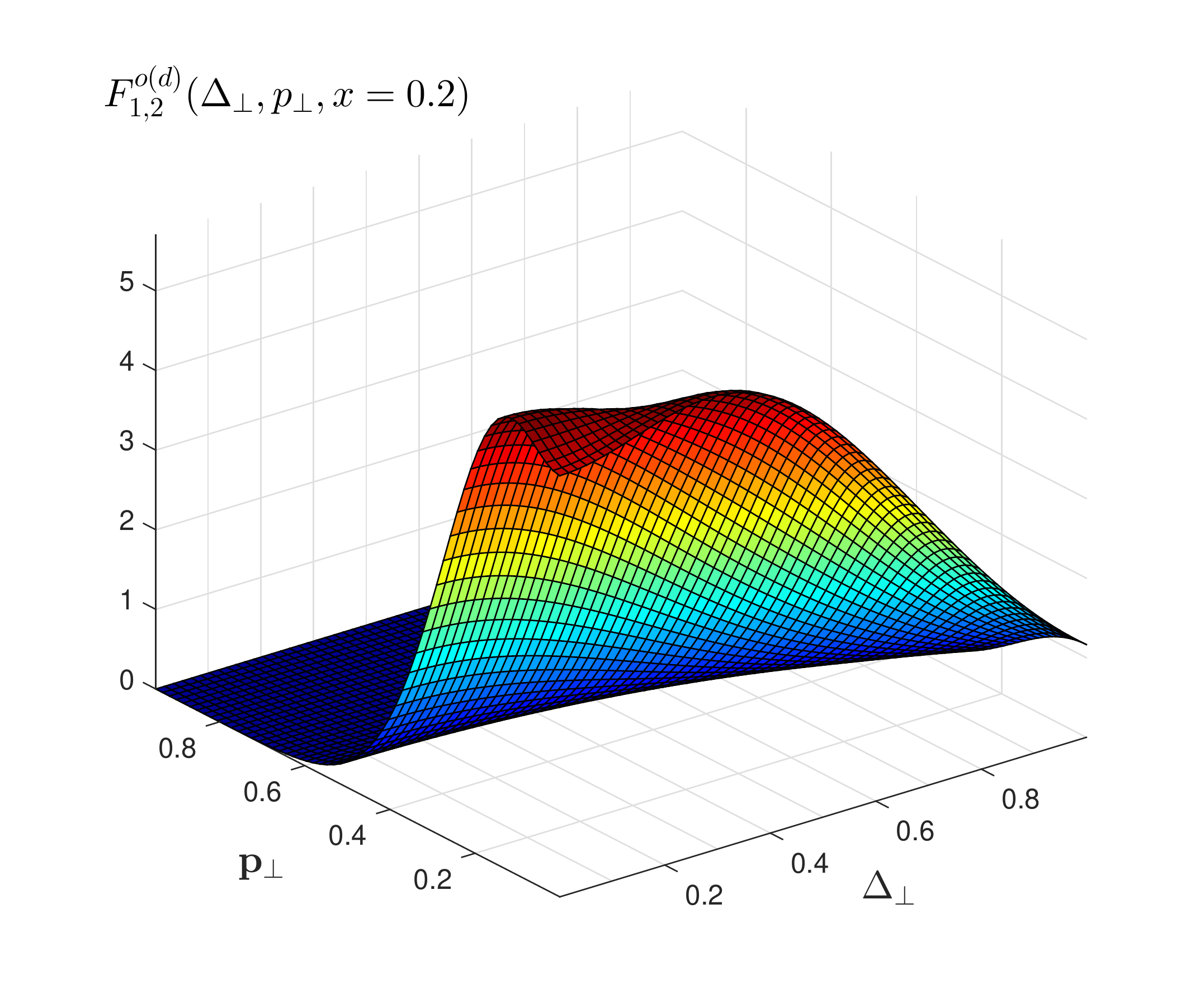} \\
\includegraphics[scale=0.38]{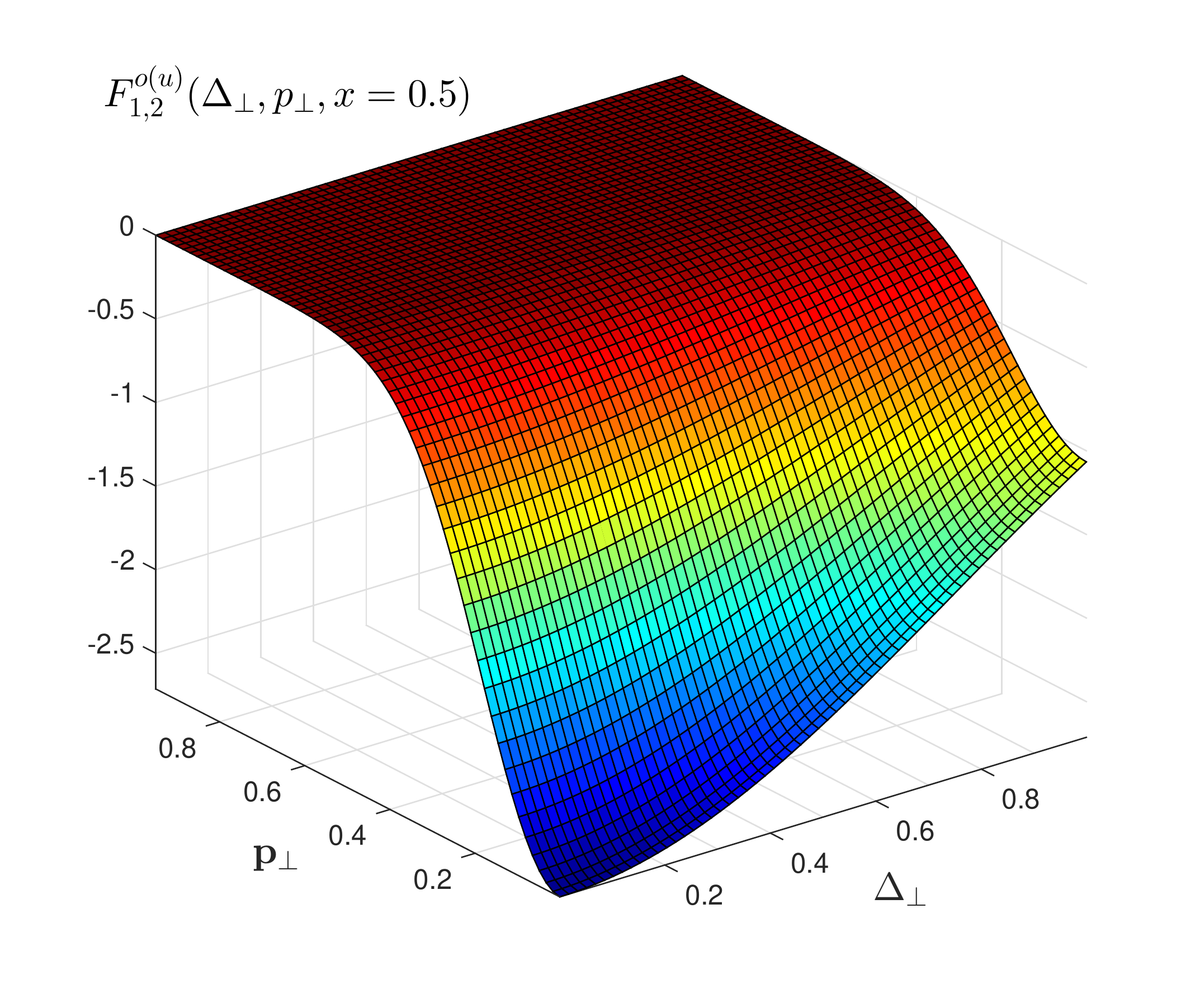}  
\includegraphics[scale=0.38]{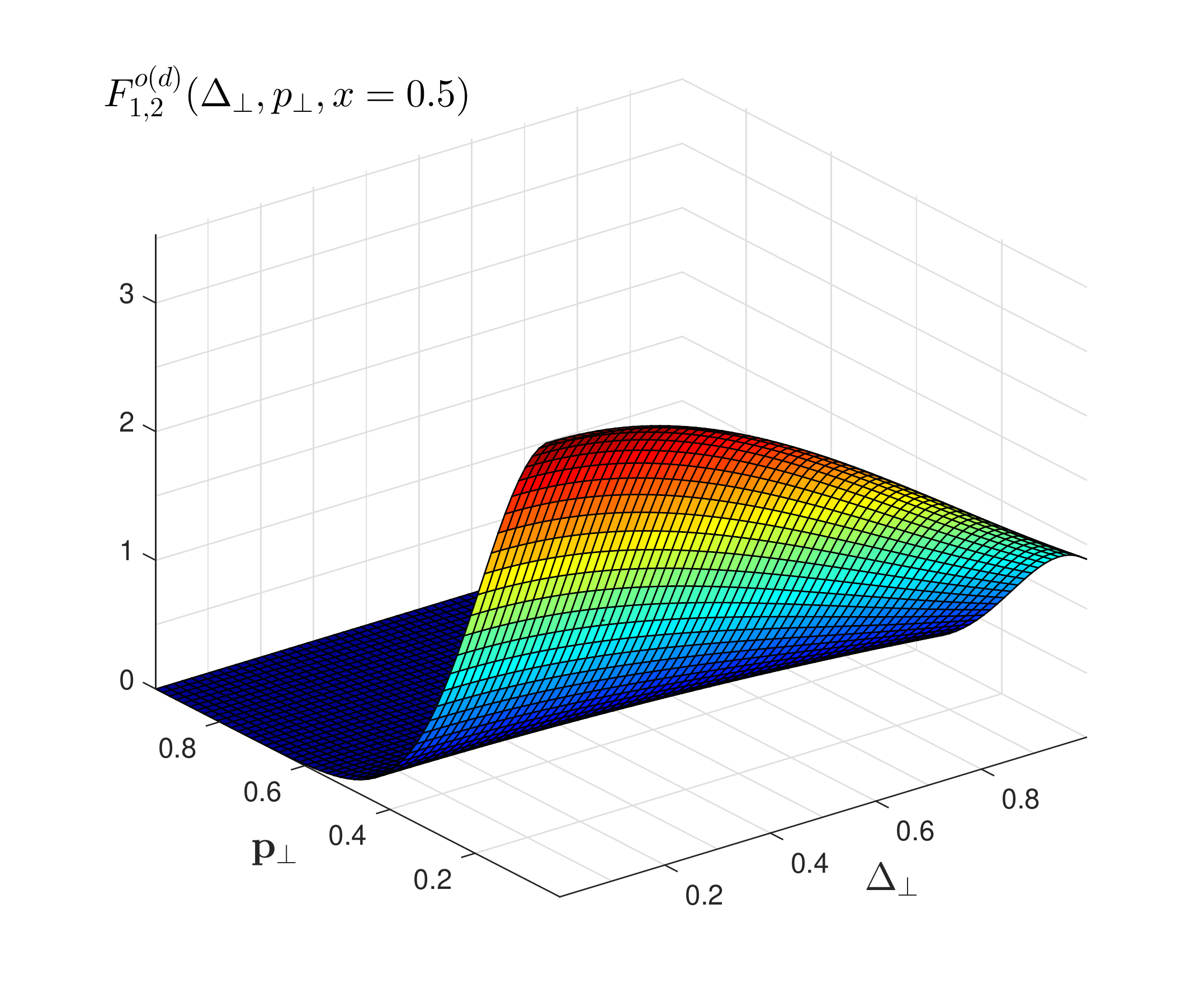}
\caption{$F^{o\nu}_{1,2}(\Delta_\perp, \bfp)$ for three different $x=0.1,0.2$ and $0.5$. The left and right columns are for $u$ and $d$ quarks respectively. \label{fig_F12o} } 
\end{figure}
\begin{figure}
\includegraphics[scale=0.38]{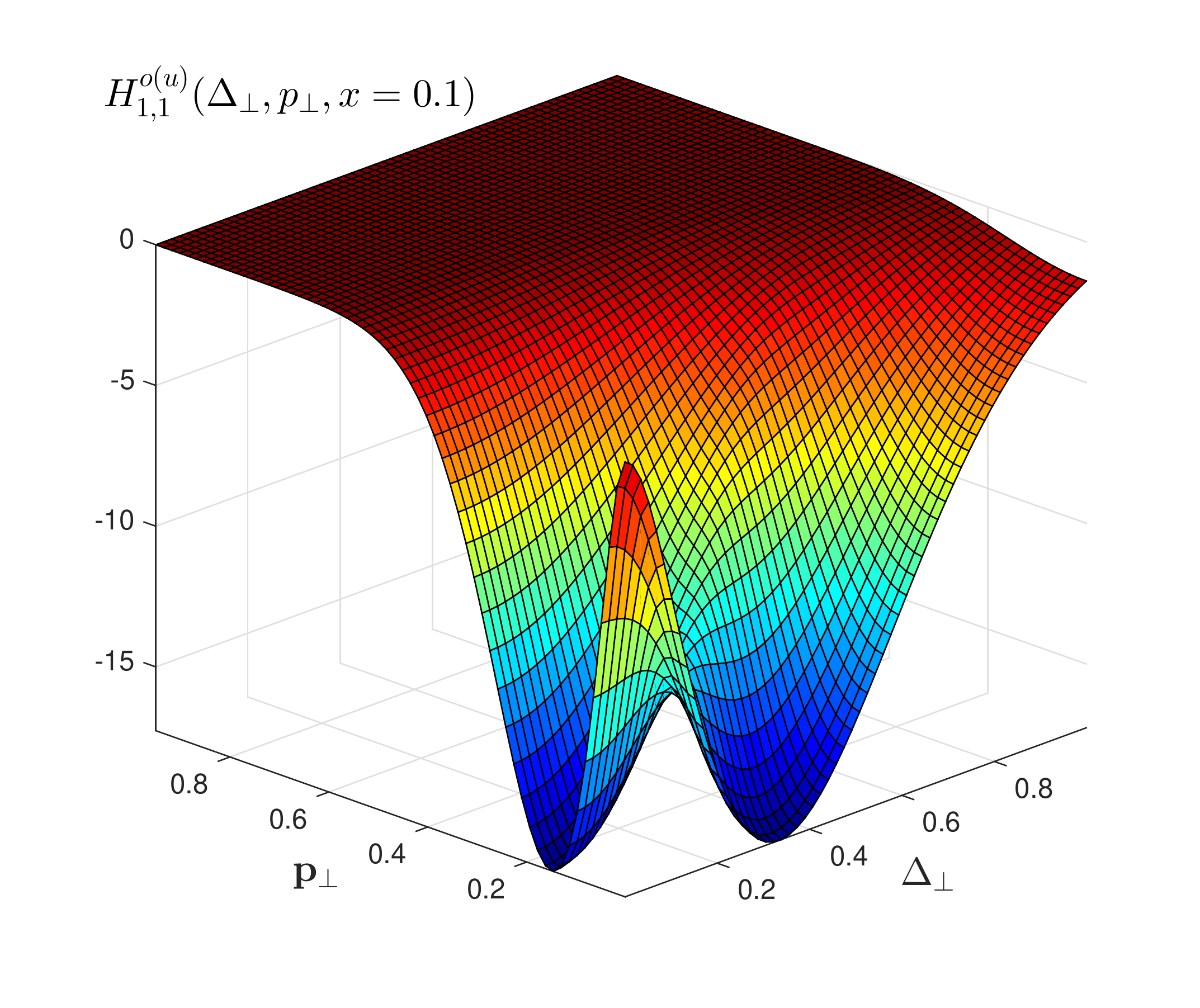} 
\includegraphics[scale=0.38]{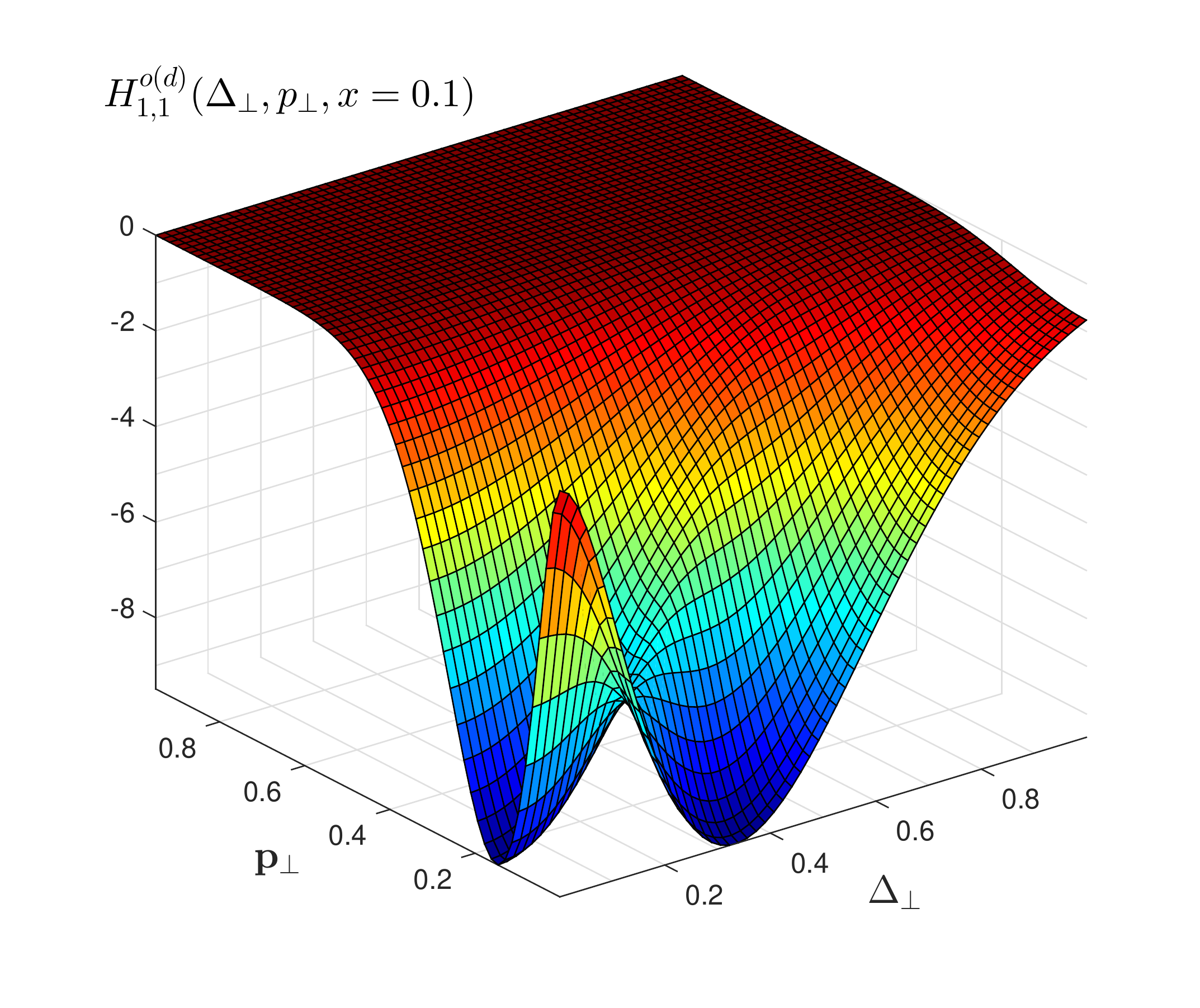} \\
\includegraphics[scale=0.38]{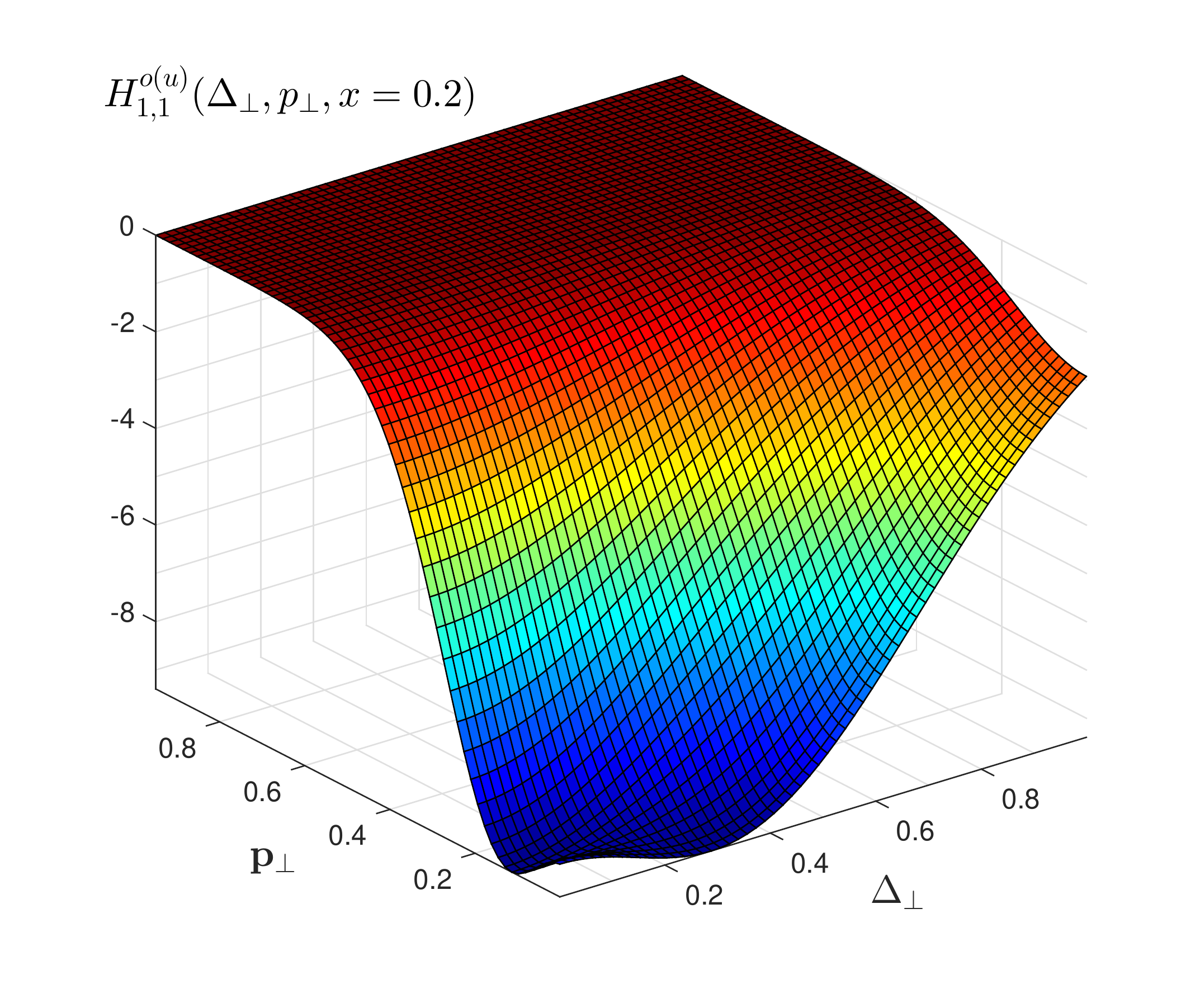}
\includegraphics[scale=0.38]{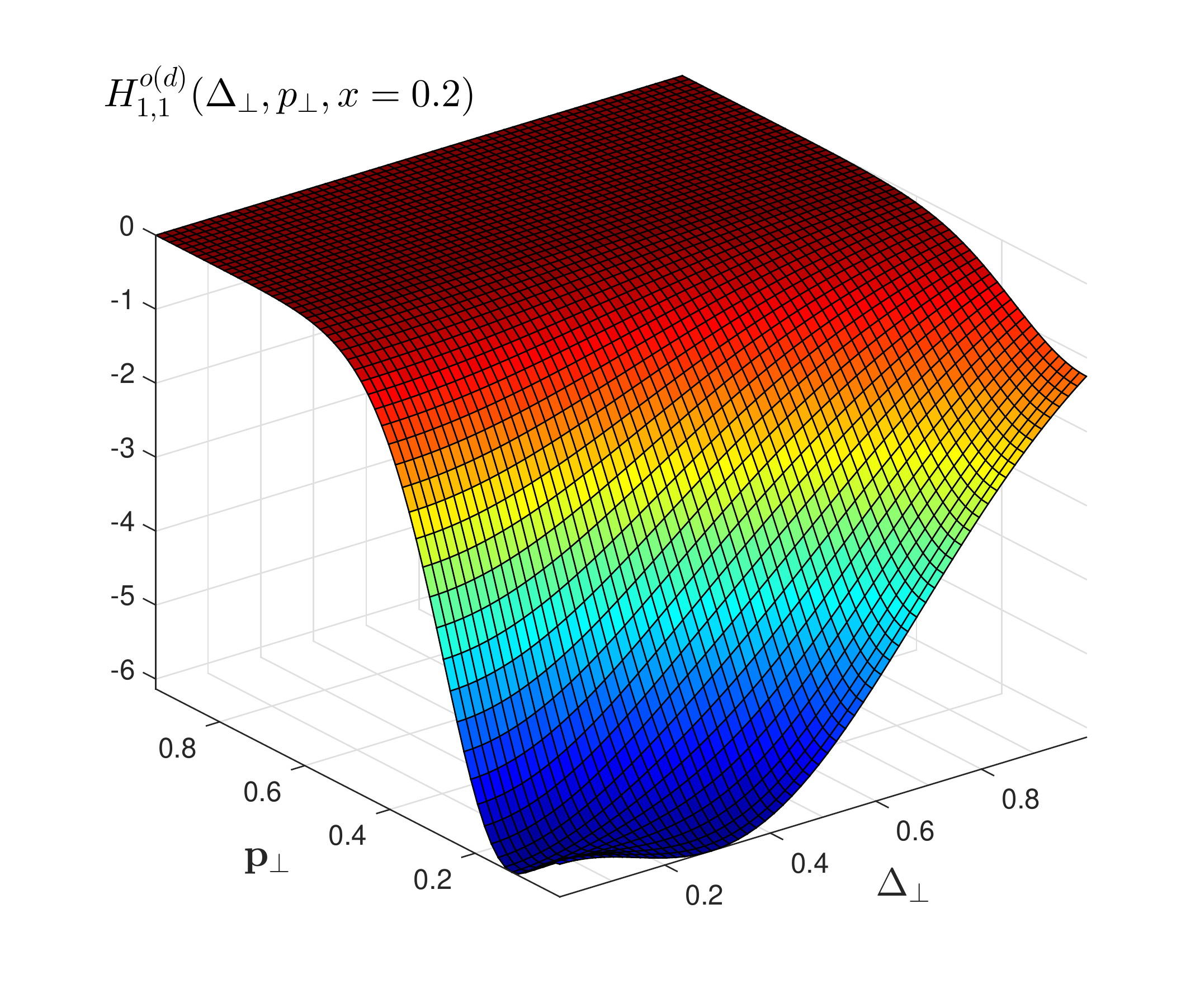} \\
\includegraphics[scale=0.38]{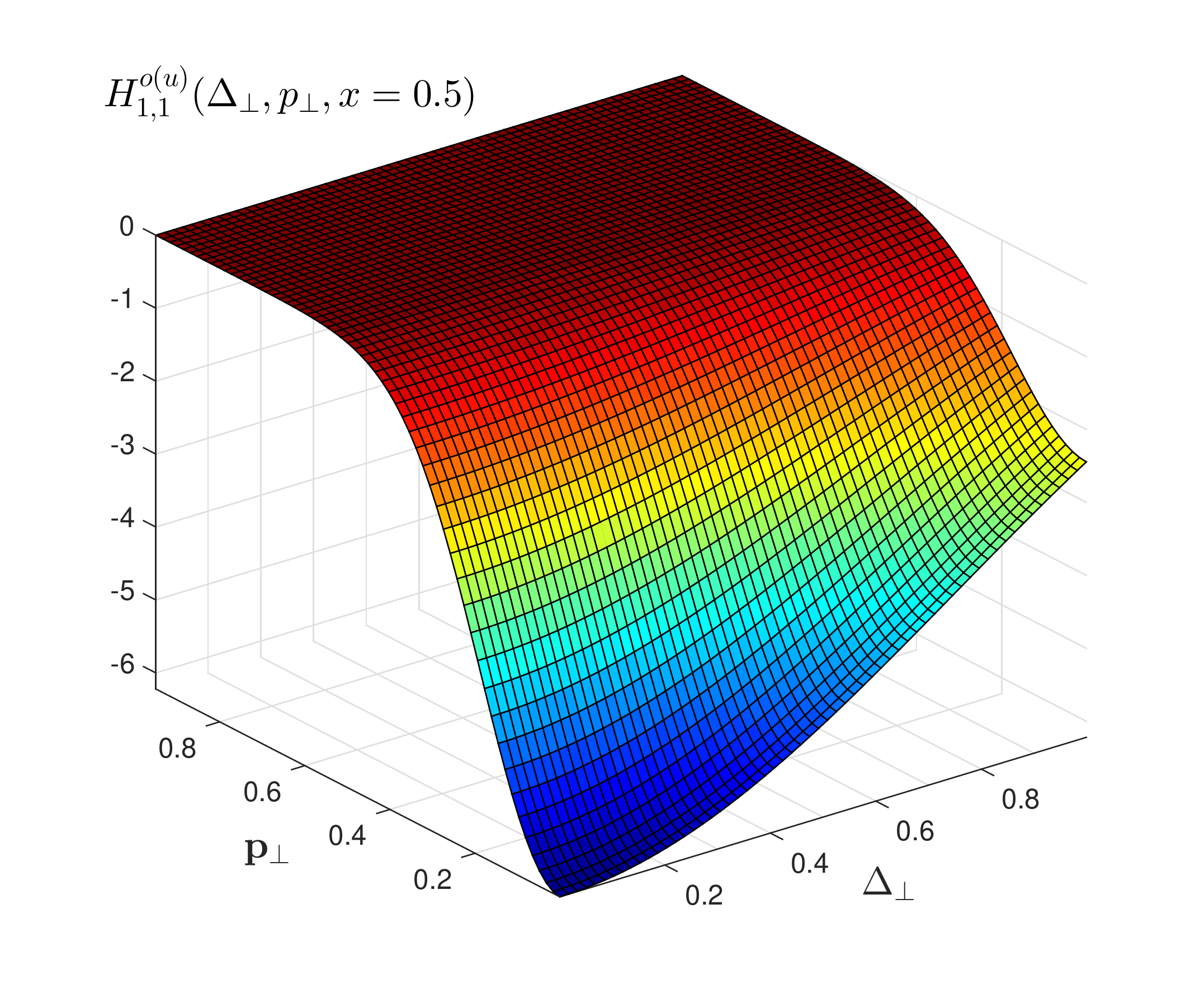}  
\includegraphics[scale=0.38]{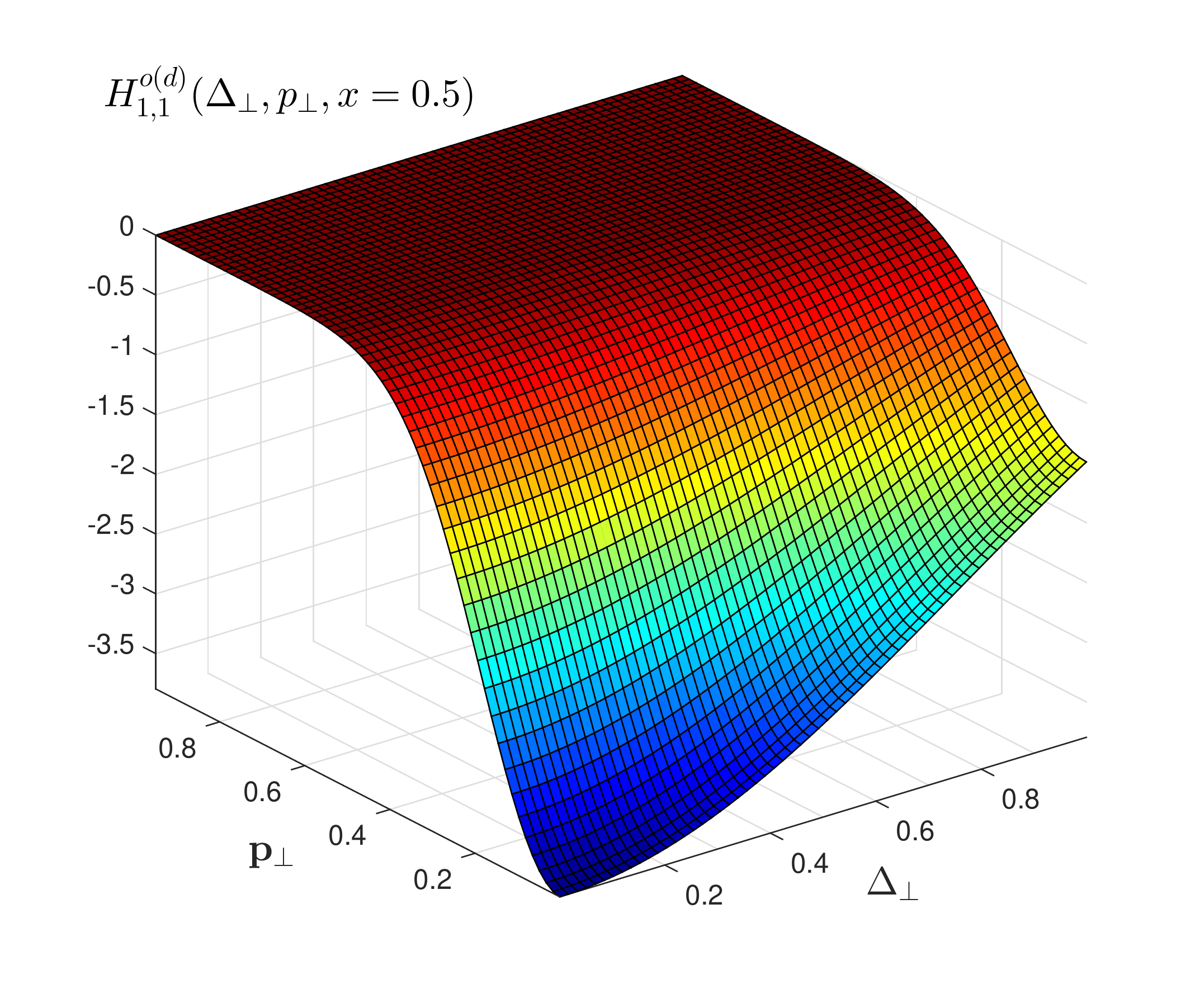} 
\caption{$H^{o\nu}_{1,1}(\Delta_\perp, \bfp)$ for three different $x=0.1,0.2$ and $0.5$. The left and right columns are for $u$ and $d$ quarks respectively. \label{fig_H11o}}
\end{figure}

\section{Results}\label{results}
In this model, using the wave functions given in Sec.(\ref{Sec_FSI}) and the proton states discussed in Sec.(\ref{model}), we calculate the GTMDs correlator defined in Eq.(\ref{GTMD_ll}) for the helicity combination $\lambda '' \lambda'$. Comparing our model results with the bilinear decomposition defined in Eq.(\ref{W_F12}) and Eq.(\ref{W_H11}), we get
\be 
F^{e\nu}_{1,2} (\Dp,\bfp,x) &=&0,\\
F^{o\nu}_{1,2} (\Dp,\bfp,x) &=& \bigg(C^2_S N^{\nu 2}_S -C^2_A \frac{1}{3}N^{\nu 2}_0 \bigg) \frac{1}{x}\bigg\{(D^\prime_1-D^\prime_2)+(D''_1-D''_2)\bigg\} \nonumber \\
&& \times \frac{1}{16\pi^3} A_1^\nu (x) A_2^\nu (x) \exp[-a(x) \tilde{\bf p}_\perp^2],\\
H^{e\nu}_{1,1} (\Dp,\bfp,x) &=&0,\\
H^{o\nu}_{1,1} (\Dp,\bfp,x) &=& \bigg(C^2_S N^{\nu 2}_S + C^2_A \big(\frac{1}{3}N^{\nu 2}_0 + \frac{2}{3}N^{\nu 2}_1\big)\bigg) \frac{1}{x}\bigg\{(D^\prime_1-D^\prime_2)+(D''_1-D''_2)\bigg\}\nonumber \\
&& \times  \frac{1}{16\pi^3} A_1^\nu (x) A_2^\nu(x) \exp[-a(x) \tilde{\bf p}_\perp^2].
\ee
In the above expressions:
\be 
D_i^\prime= -\frac{1}{2}C_F\alpha_s(\bfp^2 + B)g_i^\prime, ~&{\rm and} &~~ D_i^{\prime\prime}= -\frac{1}{2}C_F\alpha_s(\bfp^2 + B)g_i^{\prime\prime},\\
\tilde{\textbf{p}}^2_\perp &=&\bfp^2+\frac{\Dp^2}{4}(1-x)^2.
\ee
 $g_i^\prime$ and $g_i^{\prime\prime}$ are defined in the same way as Eqs.(\ref{g1},\ref{g2}) but $\bfp$ is replaced by $\bfp^\prime$ and $\bfp^{\prime\prime}$ respectively, where 
\be 
\bfp^{\prime}=\bfp-(1-x)\frac{\Dp}{2}, \quad \quad {\rm and} ~~
\bfp^{\prime\prime}=\bfp+(1-x)\frac{\Dp}{2},
\ee 
represent respectively the initial and final momentum of the struck quark.
Note that the T-even part vanishes as found before \cite{Chakrabarti:2017teq} 
and the T-odd parts became non-zero due to the incorporation of final state 
interaction.
In the final state interaction, gluon exchange strength $\frac{e_1e_2}{4\pi} \rightarrow - C_F\alpha_s$. The integrations $g_1$ and $g_2$, given in Eqs.(\ref{g1},\ref{g2}) for $i=1,2$, both have a logarithmic divergence term at the limit $m_g = 0$ ($m_g:$ mass of the exchanged gluon). The divergence terms of each integration cancel out with one another for the combination of $(g_1 - g_2)$ as well as for the combination of $(D_1-D_2)$. 
We neglect the higher order terms e.g., $D_i D_j$ ($i,j=1,2$) and present the result upto $\mathcal{O}( \alpha_s)$. The explicit results for the T-odd GTMDs in this model can be written as
\be 
F^{o\nu}_{1,2} (\Dp,\bfp,x) &=& \bigg(C^2_S N^{\nu 2}_S -C^2_A \frac{1}{3}N^{\nu 2}_0 \bigg) (-C_F \alpha_s)\frac{1}{2x}\bigg\{(\bfp^{\prime 2} + B) \frac{1}{\bfp^{\prime 2}}\ln(\frac{\bfp^{\prime 2} + B}{B})\nonumber \\
&& + (\bfp^{\prime \prime 2} + B) \frac{1}{\bfp^{\prime \prime 2}}\ln(\frac{\bfp^{\prime \prime 2} + B}{B})\bigg\} \frac{1}{16\pi^3} A_1^\nu (x) A_2^\nu (x) \exp[-a(x) \tilde{\bf p}_\perp^2], \label{F12_LFQDM}\\
H^{o\nu}_{1,1} (\Dp,\bfp,x) &=& \bigg(C^2_S N^{\nu 2}_S + C^2_A \big(\frac{1}{3}N^{\nu 2}_0 + \frac{2}{3}N^{\nu 2}_1\big)\bigg) (-C_F \alpha_s)\frac{1}{2x}\bigg\{(\bfp^{\prime 2} + B) \frac{1}{\bfp^{\prime 2}}\ln(\frac{\bfp^{\prime 2} + B}{B})\nonumber \\
&& + (\bfp^{\prime \prime 2} + B) \frac{1}{\bfp^{\prime\prime 2}}\ln(\frac{\bfp^{\prime \prime 2} + B}{B})\bigg\}  \frac{1}{16\pi^3} A_1^\nu(x) A_2^\nu(x) \exp[-a(x) \tilde{\bf p}_\perp^2]\label{H11_LFQDM}.
\ee

From the above equation, one can find out a model dependent relation between these two T-odd TMDs as
\be 
\frac{F^{o\nu}_{1,2} (\Dp,\bfp,x)}{H^{o\nu}_{1,1} (\Dp,\bfp,x)}= \frac{\bigg(C^2_S N^{\nu 2}_S -C^2_A \frac{1}{3}N^{\nu 2}_0 \bigg)}{\bigg(C^2_S N^{\nu 2}_S + C^2_A \big(\frac{1}{3}N^{\nu 2}_0 + \frac{2}{3}N^{\nu 2}_1\big)\bigg)} = R^\nu. \label{Rel_GTMD}
\ee 
In this model $R^u=0.4363  $ for u quark and $R^d= -0.9290 $ for d quark. The Sivers and Boer-Mulders Wigner distributions are also related with the same constant $R^\nu$ ( as they are defined as the  Fourier Transform of these GTMDs).   

Our model result for T-odd GTMDs $F^o_{1,2} (\Dp,\bfp,x)$ is shown in 
Fig.(\ref{fig_F12o}) in the $\bfp$ and $\Dp$ plane for different values of $x$. 
The three rows are for three different values of $x=0.1,0.2$ and $0.5$ and the 
first and second columns are for $u$ and $d$ quarks respectively. One can notice that the $F^o_{1,2}$ is mostly negative for for $u$ quark and positive side for $d$ quark. The distributions become flat near $(0,0)$ with the increasing value of $x$.

Fig.\ref{fig_H11o} shows the model result for $H^o_{1,1} (\Dp,\bfp,x)$ for $u$ and $d$ quarks with a different values of $x=0.1,0.2$ and $0.5$. Here, distributions does not change sign for the change in flavor unlike   $F^o_{1,2} (\Dp,\bfp,x)$.

In Eqs.(\ref{F12_LFQDM},\ref{H11_LFQDM}), the values of the constants $C^2_S$ for scalar diquark state and $C^2_A$($A=V,VV$ for the diquark with configuration $ud$ and $uu$) for axial-vector are given in \cite{Maji:2016yqo} for this model. We also refer \cite{Maji:2016yqo} for the values of the normalization constants $N^\nu_S,N^\nu_0,N^\nu_1$ with flavor $\nu=u,d$. 
For the numerical calculation we take proton mass $M=0.94~ GeV$, diquark mass and quark mass are $m_D=0.894~GeV$($m^2_D=0.8~GeV^2$) and $m_q=0.055 GeV$($m^2_q=0.003~GeV^2$). The color factor $C_F=4/3$ and the strong coupling is taken as $\alpha_s=0.3$. The non zero quark mass are used as a regulator such as $(m_q+m_D) > M$ to have free bound state proton. We use the AdS/QCD scale parameter $\kappa =0.4~GeV$ as determined in \cite{Chakrabarti:2013gra}. 

\begin{figure}[htbp]
\begin{minipage}[c]{0.98\textwidth}
\small{(a)}\includegraphics[width=6.5cm,clip]{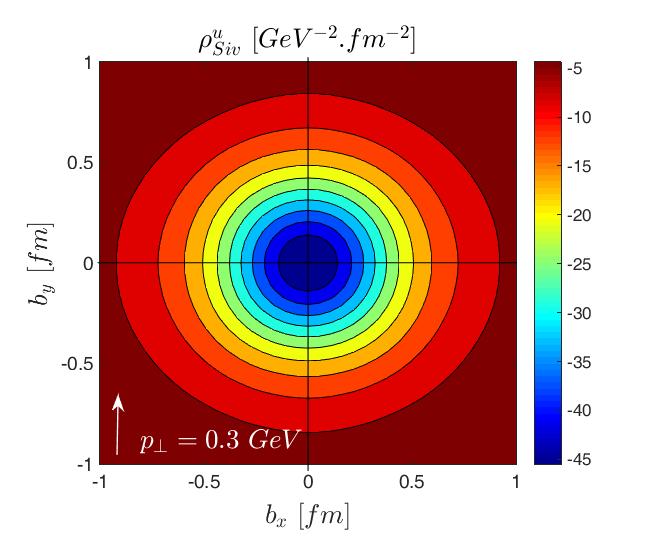}
\small{(b)}\includegraphics[width=6.5cm,clip]{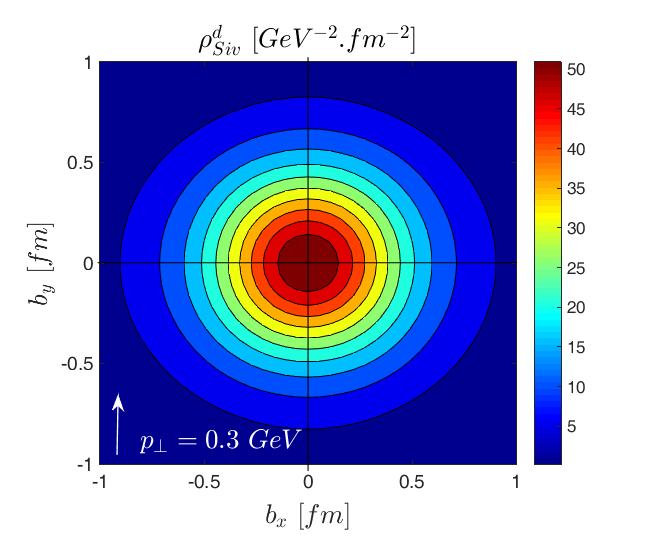}
\end{minipage}
\begin{minipage}[c]{0.98\textwidth}
\small{(c)}\includegraphics[width=6.5cm,clip]{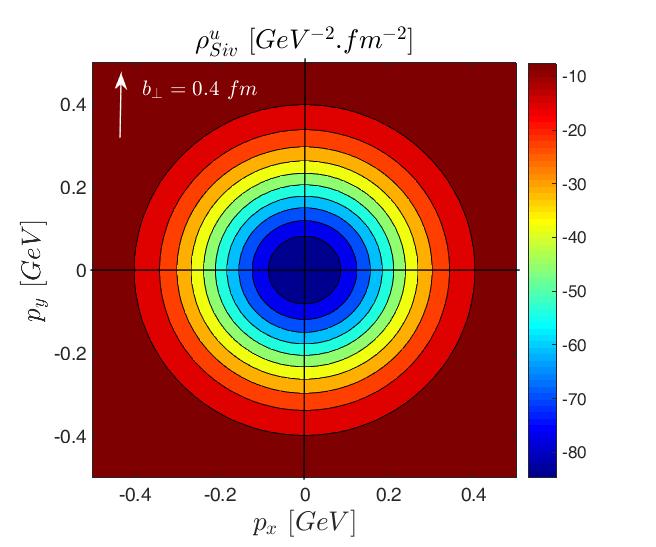}
\small{(d)}\includegraphics[width=6.5cm,clip]{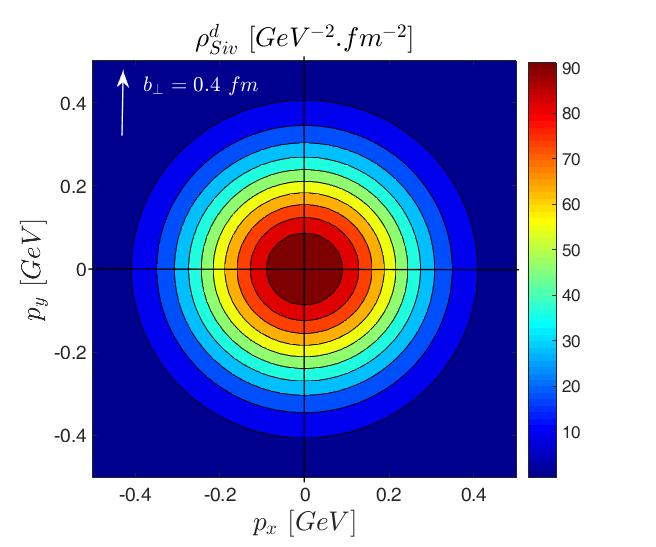} 
\end{minipage}
\caption{\label{fig_Siv} $x$ integrated Sivers Wigner distribution $\rho^{\nu}_{Siv}((\bfb,\bfp))$: in the impact-parameter plane(a,b) for $\bfp=0.3~GeV$ along $\hat{y}$ and in the transverse momentum plane (c,d) with $\bfb=0.4~fm$ along $\hat{y}$. The left and right columns are for $u$ and $d$ quarks respectively.} 
\end{figure} 
Using the model result from Eq.(\ref{F12_LFQDM}), we take a Fourier transformation of $F^o_{1,2}(\Dp,\bfp,x)$(Eq.(\ref{RSiv})) and get the Sivers Wigner distribution  $\rho^\nu_{Siv}(\bfb,\bfp,x)$. The model result for $x$ integrated Sivers Wigner distribution in the transverse $\bfb$ plane is shown in Fig.\ref{fig_Siv}(a,b) for $u$ and $d$ quarks respectively.  $\rho^\nu_{Siv}(\bfb,\bfp)$ is also shown in the transverse momentum plane in Fig.\ref{fig_Siv}(c,d). The distribution is axially symmetric in both impact- parameter plane as well as in the transverse momentum plane for $u$ and $d$ quarks.      
  
\begin{figure}[htbp]
\begin{minipage}[c]{0.98\textwidth}
\small{(a)}\includegraphics[width=6.5cm,clip]{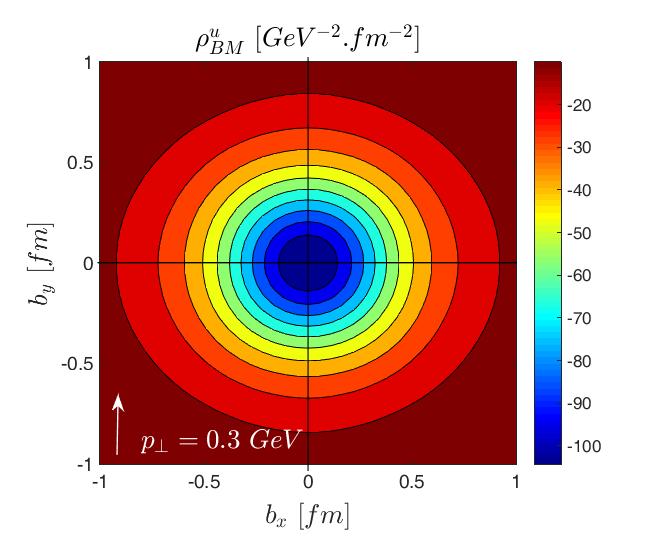}
\small{(b)}\includegraphics[width=6.5cm,clip]{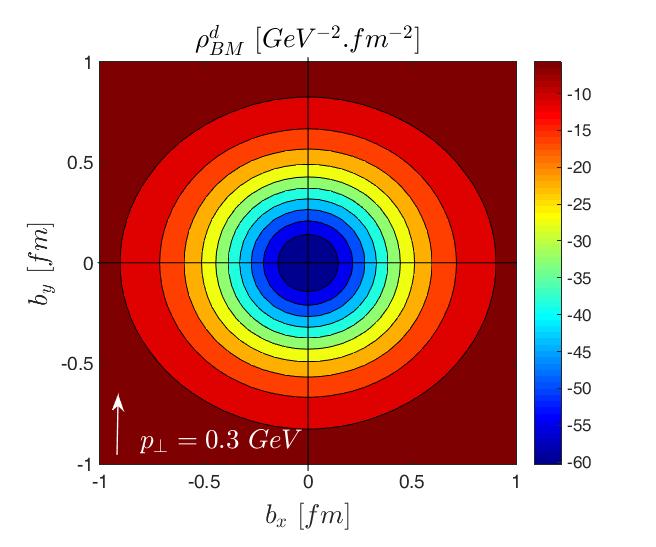}
\end{minipage}
\begin{minipage}[c]{0.98\textwidth}
\small{(c)}\includegraphics[width=6.5cm,clip]{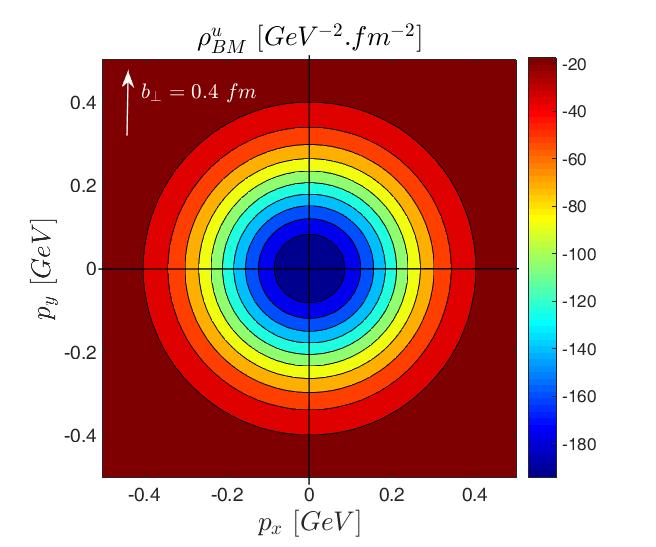}
\small{(d)}\includegraphics[width=6.5cm,clip]{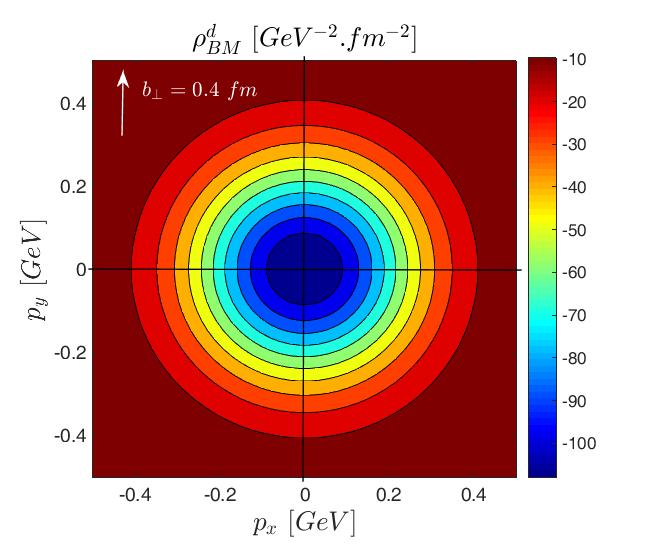} 
\end{minipage}
\caption{\label{fig_BM} $x$ integrated Boer-Mulders Wigner distribution $\rho_{BM}(\bfb,\bfp)$: in the impact-parameter plane (a,b) for $\bfp=0.3~GeV$ along $\hat{y}$ and in the transverse momentum plane (c,d) with $\bfb=0.4~fm$ along $\hat{y}$. The left and right columns are for $u$ and $d$ quarks respectively.} 
\end{figure} 
Model results for the $x$ integrated Boer-Mulders WD in the  impact- parameter plane as well as in the transverse momentum plane are shown in Fig.(\ref{fig_BM}). The distributions are axially symmetric for both $u$ and $d$ quarks in both the $\bfb$ and $\bfp$ planes. 

\begin{figure}[htbp]
\begin{minipage}[c]{0.98\textwidth}
\small{(a)}\includegraphics[width=6.5cm,clip]{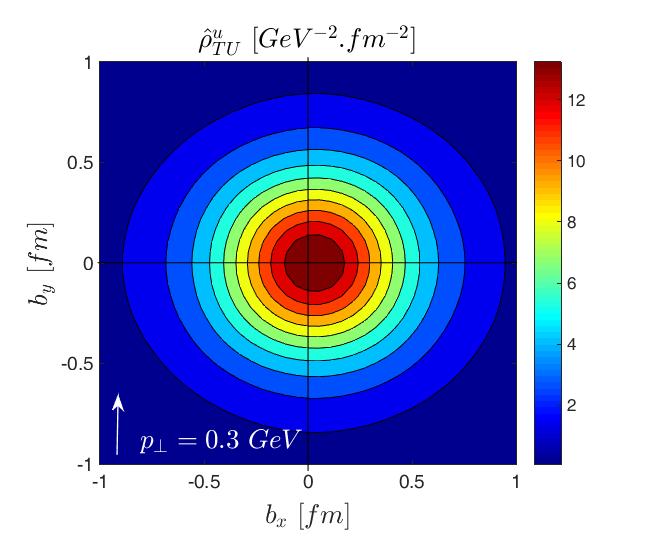}
\small{(b)}\includegraphics[width=6.5cm,clip]{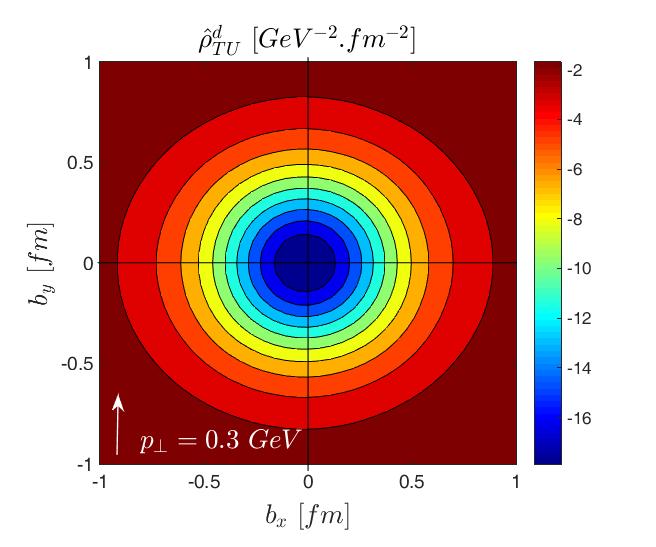}
\end{minipage}
\begin{minipage}[c]{0.98\textwidth}
\small{(c)}\includegraphics[width=6.5cm,clip]{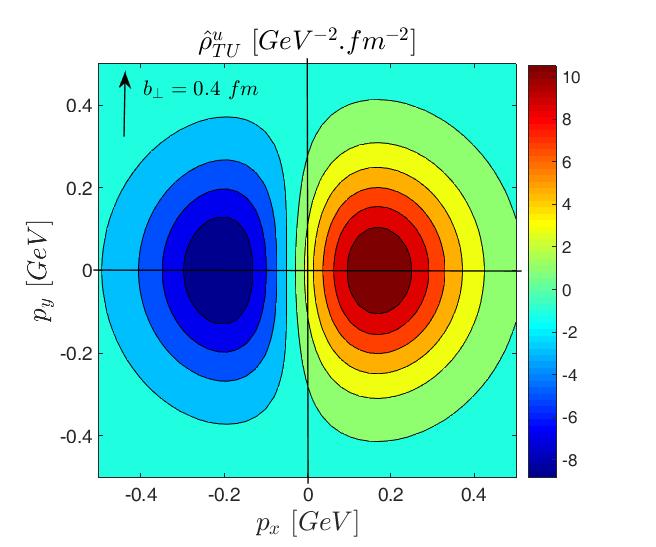}
\small{(d)}\includegraphics[width=6.5cm,clip]{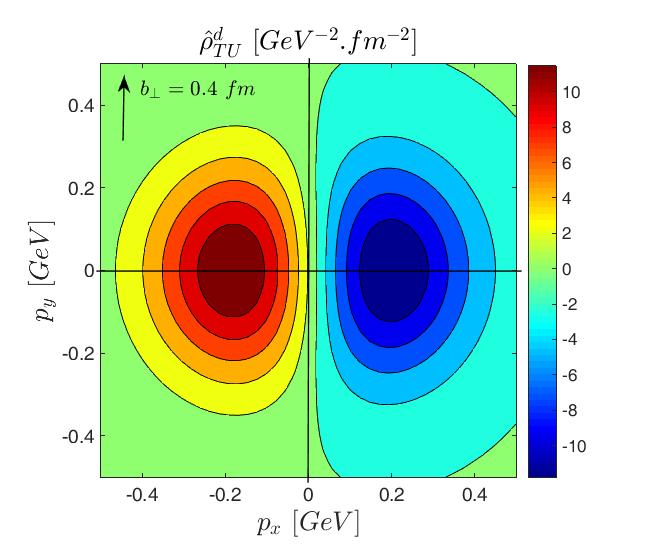} 
\end{minipage}
\caption{\label{fig_RTUSiv} The modified WD $\hat{\rho}^\nu_{TU}(\bfb,\bfp)$ in the impact-parameter plane (a,b) for $\bfp=0.3~GeV$ along $\hat{y}$ and in the transverse momentum plane (c,d)for $\bfb=0.4~fm$ along $\hat{y}$. The left and right columns are for $u$ and $d$ quarks respectively.}
\end{figure} 
Using Eq.(\ref{RUTSiv}), we present our model result for the modified Wigner Distribution $\hat{\rho}^\nu_{TU}(\bfb,\bfp)$ in Fig.\ref{fig_RTUSiv} where the proton is transversely polarised along $\hat{x}$ corresponding to the superscript $i=1$. In Fig.\ref{fig_RTUSiv}, first row is for impact-parameter plane and the second row is for transverse momentum plane. According to Eq.(\ref{RUTSiv}), $\rho_{Siv}$ has the coefficient $-p_2/M$ for $i=1$ and we take $\bfp=0.3~GeV$ along $\hat{y}$. If one considers $\bfp$ along $\hat{x}$, for $i=1$, the contribution from the FSI term  will vanish. In case of transverse momentum plane, $\hat{\rho}^\nu_{TU}(\bfb,\bfp)$ shows a dipolar nature because of presence of the factor $-p_2/M$ in the coefficient of $\rho_{siv}$.  Since the T-odd GTMDs $F_{1,2}^o$ has opposite sign for $u$ and $d$ quarks (see Fig.(\ref{fig_F12o})), the polarity of the $d$ quark distribution is opposite to the $u$ quark. Here, if one choose $i=2$ the axis of the dipolar distribution will change to the $y$ axis. It is to be noted that unlike in \cite{Lorce:2015sqe}, here we have not done a multipole decomposition of the phase space distributions.

\begin{figure}[htbp]
\begin{minipage}[c]{0.98\textwidth}
\small{(a)}\includegraphics[width=6.5cm,clip]{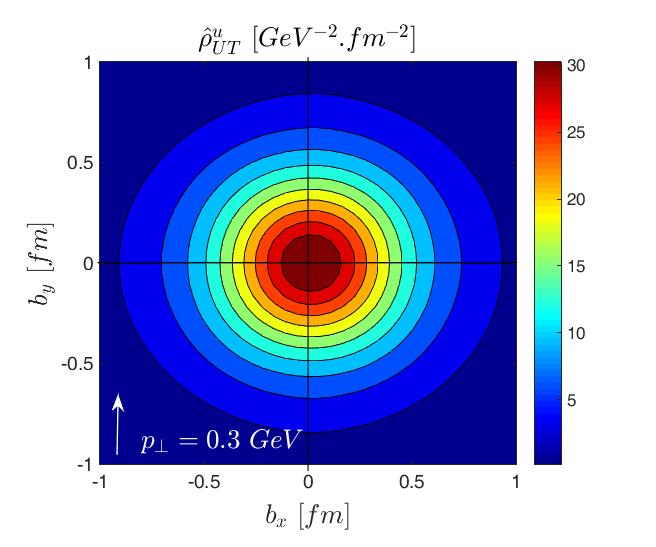}
\small{(b)}\includegraphics[width=6.5cm,clip]{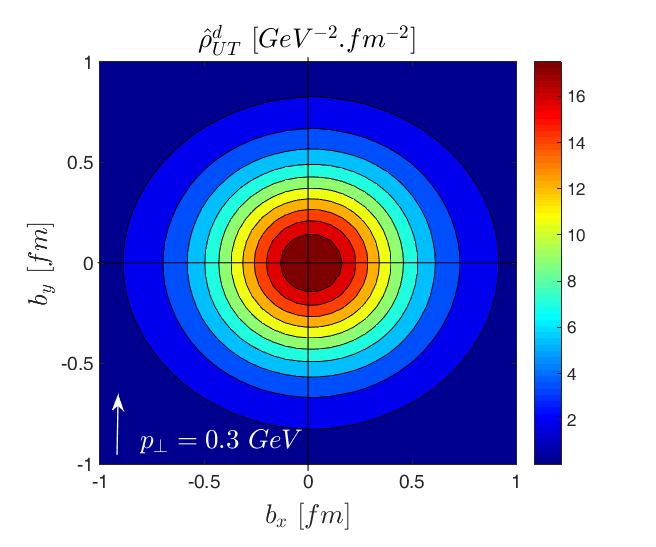}
\end{minipage}
\begin{minipage}[c]{0.98\textwidth}
\small{(c)}\includegraphics[width=6.5cm,clip]{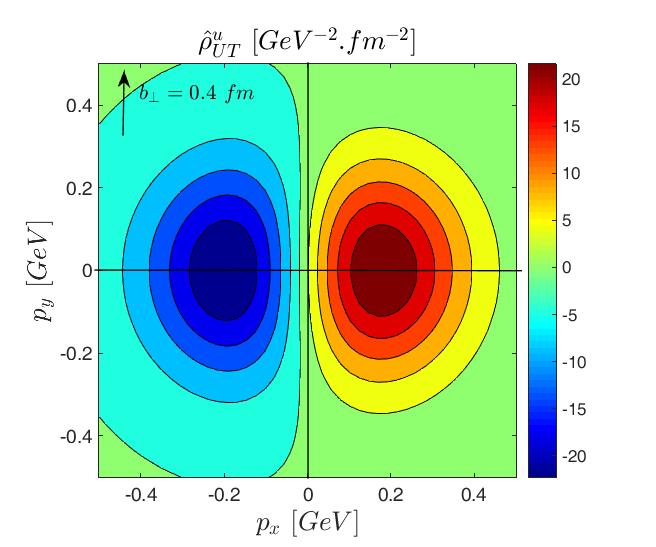}
\small{(d)}\includegraphics[width=6.5cm,clip]{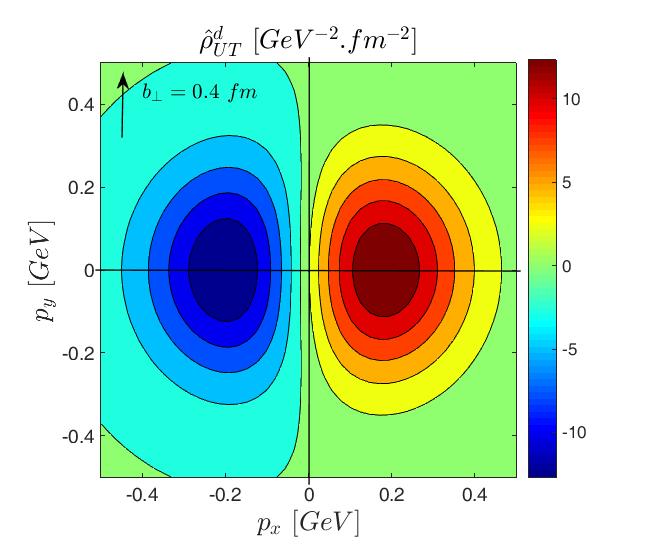} 
\end{minipage}
\caption{\label{fig_RUTBM} The modified WD $\hat{\rho}^\nu_{UT}(\bfb,\bfp)$ in the impact-parameter plane (a,b) for $\bfp=0.3~GeV$ along $\hat{y}$ and in the transverse momentum plane (c,d) for $\bfb=0.4~fm$ along $\hat{y}$. The left and right columns are for $u$ and $d$ quarks respectively.}
\end{figure} 

Fig.\ref{fig_RUTBM} represents the modified $\hat{\rho}^\nu_{UT}(\bfb,\bfp)$ in the impact-parameter plane, $(a,b)$ with $\bfp=0.3~GeV$ along $\hat{y}$ and in the transverse momentum plane  $(c,d)$ with $\bfb=0.4~fm$ along $\hat{y}$ for both the quarks. The quark polarisation is taken to be along $x$ axis, i.e, for $j=1$ in Eq.(\ref{RUTBM}). We observe an axially symmetric distribution in the impact-parameter plane and a dipolar distribution in the transverse momentum plane. Unlike $\hat{\rho}^\nu_{TU}(\bfb,\bfp)$,  the polarity of $\hat{\rho}^\nu_{UT}(\bfb,\bfp)$ is same for both $u$ and $d$ quarks. This is due to the fact that the T-odd GTMD $ H_{1,1}^o$ has same sign for both $u$ and $d$ quarks(see Fig.(\ref{fig_H11o})).

We refer \cite{Chakrabarti:2017teq} for the model results of the Wigner distributions  without incorporating the FSI effects  e.g., Wigner distributions $\rho^\nu_{TU}(\bfb,\bfp,x)$ and $\rho^\nu_{UT}(\bfb,\bfp)$. 

\section{Final State Interaction Function}\label{fsi}
It is known that in spectator type models, T-odd TMDs  can be factored out into a term coming from the FSI (Wilson line) at the level of one gluon exchange, and a part independent of the FSI \citep{Bacchetta:2008af}. This has been shown for the Sivers function with scalar diquarks in \citep{Burkardt:2003je,Lu:2006kt} and with axial vector diquarks in \citep{Bacchetta:2008af}, also for Boer-Mulders function. In \citep{Bacchetta:2008af} several spectator type models were used to calculate the TMDs, and the FSI function was found to depend on the form factor of the specific model. Burkardt in \citep{Burkardt:2002ks}  in a spectator model calculation showed a connection between the average transverse momentum of the quarks generated by Sivers function and the distortion in impact-parameter space related to the GPD $E_q$, through a term generated by the FSI  called lensing function.  The lensing function is related to the transverse impulse on the quark and is a model dependent quantity. In general, such relation between the GPDs and TMDs cannot be obtained in a model independent way.  In our model, we are able to  factor out a FSI function in  the TMD limit:
\be
G(x,\bfp,\hat{\bf{p}}_\perp)= - C_F \alpha_S (\hat{\bf{p}}_\perp^2+B) \frac{1}{\hat{\bf{p}}_\perp^2}\ln\bigg(\frac{\hat{\bf{p}}_\perp^2+B}{\hat{\bf{p}}_\perp^2}\bigg) \delta^2(\hat{\bf{p}}_\perp-\bfp)
\ee
FSI function for the GTMDs can be written as
\be
G(x,\bfp,\hat{\bf{p}}_\perp,\Delta_\perp)&=& - C_F \alpha_S \bigg[(\hat{\bf{p}}_\perp'^2+B) \frac{1}{\hat{\bf{p}}_\perp'^2}\ln\bigg(\frac{\hat{\bf{p}}_\perp'^2+B}{\hat{\bf{p}}_\perp'^2}\bigg)\nonumber \\
&& \hspace{2cm}+ (\hat{\bf{p}}_\perp''^2+B) \frac{1}{\hat{\bf{p}}_\perp''^2}\ln\bigg(\frac{\hat{\bf{p}}_\perp''^2+B}{\hat{\bf{p}}_\perp''^2}\bigg) \bigg] \delta^2(\hat{\bf{p}}_\perp-\bfp)
\ee
These are the same for the Sivers and Boer-Mulders GTMDs.  In the above equation 
\be 
\hat{{\bf p}}''_\perp = \hat{{\bf p}}_\perp + \frac{\Dp}{2}(1-x)^2, ~\hspace{1cm}~ \hat{{\bf p}}'_\perp = \hat{{\bf p}}_\perp - \frac{\Dp}{2}(1-x)^2
\ee
where, $\hat{{\bf p}}_\perp$ represents the free momentum in the one gluon exchange loop and eventually integrated out in the correlator calculation. 
At the TMD limit $\Dp=0$, final state interaction function for GTMD reduces to the TMD FSI function.

\section{Summary and Conclusion}\label{summary}
Wigner distributions and the GTMDs have drawn a lot of attention in recent 
years as various experiments are probing the internal structure of proton. 
T-even GTMDs and TMDs are studied in several models but T-odd GTMDs are not 
investigated much.   Final state interaction introduces a complex phase in the 
amplitude which generates the spin asymmetries associated with the T-odd TMDs. 
In this work, we have incorporated the effect of the FSI in the wave functions 
for a model study of the T-odd GTMDs and Wigner distributions.

The T-odd GTMDs $F_{1,2}^o$ and $H_{1,1}^o$ reduce to Sivers and Boer-Mulders 
functions in the TMD limit $(\Delta_\perp=0)$.  
It is interesting to note that both these T-odd GTMDs have exactly same 
functional dependence on $\Delta_\perp,p_\perp$ and $x$, they only differ in 
the overall normalization factors which also depend on the quark flavor in our 
model. Since the sign of the prefactor in $F_{1,2}^o(\Delta_\perp,p_\perp,x)$ 
is positive for $u$ and negative for $d$ quark, the GTMD 
$F_{1,2}^o(\Delta_\perp,p_\perp,x)$ is found be negative for $u$ quark while it 
is positive for $d$ quark. But, in 
$H_{1,1}^o (\Delta_\perp,p_\perp,x)$, the prefactor is always positive and 
hence the GTMD is negative for both $u$ and $d$ quarks. The 
Fourier transform of the GTMDs when integrated over $x$ give the 
Sivers Wigner distributions $\rho^\nu_{siv}$ and BM Wigner distributions 
$\rho^\nu_{BM}$ which are found to be axially symmetric in both $p_\perp$ and $b_\perp$ plane. 

The Wigner distributions get modified due to contributions from the T-odd GTMDs. The  FSI contributions to the  Wigner distributions are proportional to the transverse momentum. As a result, the distributions when plotted in the 
momentum plane, are bipolar in nature. But it is interesting to note that in the $b_\perp$ plane, the distributions with T-even GTMDs were bipolar, but 
the contributions from T-odd part wash out that bipolar behaviour. The asymmetries in the modified distributions  $\hat{\rho}_{TU}$ and 
$\hat{\rho}_{UT}$ in the $p_\perp$ plane are very mild as the centre of the 
distributions are slightly shifted from the centre $(0,0)$.
We have also shown that the FSI contribution can be factored out  for the GTMDs in our model, as observed in  other spectator type models.

\section {Acknowledgement}
We thank Cedric Lorce for fruitful discussions. The work of T.M. is supported by National Natural Science Foundation of China (NSFC) through Grant No. 11875112.

\appendix*
\section{T-odd GTMDs} \label{appA}
For completeness, we list our model results for other (related) T-odd GTMDs here.
\be 
F^{o \nu}_{1,1} (\Dp,\bfp,x) &=& \bigg(C^2_S N^{\nu 2}_S + C^2_A \big(\frac{1}{3}N^{\nu 2}_0 + \frac{2}{3}N^{\nu 2}_1\big)\bigg) \frac{1}{16 \pi^3}\bigg[ (D^\prime_1-D''_1)|A_1|^2 \nonumber \\
 && + \frac{1}{x^2 M^2} \bigg(\bfp^2 - \frac{\Delta_\perp^2}{4}(1-x)^2\bigg) (D'_2-D''_2)|A^\nu_2|^2 \bigg] \exp[-a(x) \tilde{\bf p}_\perp^2]\\
F^{o \nu}_{1,3} (\Dp,\bfp,x) &=& \frac{1}{2} F^o_{1,1}+\bigg(C^2_S N^{\nu 2}_S + C^2_A \frac{1}{3}N^{\nu 2}_0 \bigg) \frac{(1-x)}{2x} (D'_1+D'_2-D''_1-D''_2)\nonumber \\
&& \times \frac{1}{16 \pi^3} A^\nu_1(x)A^\nu_2(x) \exp[-a(x) \tilde{\bf p}_\perp^2]\\
H^{o \nu}_{1,2} (\Dp,\bfp,x) &=& \bigg(C^2_S N^{\nu 2}_S + C^2_A \big(\frac{1}{3}N^{\nu 2}_0 + \frac{2}{3}N^{\nu 2}_1\big)\bigg) \frac{(1-x)}{x}(D'_1 + D'_2 - D''_1 - D''_2) \nonumber \\
&& \times \frac{1}{16 \pi^3} A^\nu_1(x)A^\nu_2(x) \exp[-a(x) \tilde{\bf p}_\perp^2]
\ee
All the above T-odd distributions have logarithmic divergence as  the divergent terms cancel out for the combinations $(D'_1-D'_2)$ and $(D''_1-D''_2)$. This divergence comes from the gluon mass being set to zero, and a regulator is needed for numerical evaluations. At $\Delta_\perp = 0$ \cite{Meissner:2009ww}, these T-odd GTMDs do not have any TMD limit.

\bibliographystyle{apsrev}

 \end{document}